\documentclass[journal]{IEEEtran} 

\usepackage{amsmath,amsthm}
\usepackage{dsfont}
\usepackage{algorithm, algorithmic}
\usepackage{graphicx}

\newtheorem{lemma}{Lemma}

\newtheorem{Theorem}{Theorem}

\newtheorem{defi}{Definition}
\newtheorem{prop}{Proposition}
\newtheorem{corol}{Corollary}

\newtheorem{con}{Conclusion}
\newtheorem{Example}{Example}

\def\rank{{\rm rank}}
\def\row{{\rm row}}
\def\col{{\rm col}}
\def\dim{{\rm dim}}
\def\Pr{{\rm Pr}}
\def\rowspan{{\rm rowspan}}

\def\argmin{{\rm argmin}}
\def\argmax{{\rm argmax}}

\def\Real{\mathop{\hbox{\mit I\kern-.2em R}}\nolimits}
\def\Zeal{\mathop{\hbox{\mit Z\kern-.29em Z}}\nolimits}
\def\be{\begin{equation}}
\def\ee{\end{equation}}

\def\ba{\begin{eqnarray*}}
\def\ea{\end{eqnarray*}}

\def\bit{\bibitem}

\input{epsf}

\ifCLASSINFOpdf
\else
\fi

\begin{document}
%
\title{Performance Bounds on a Wiretap Network with Arbitrary Wiretap Sets}
%
%
%

\author{Fan Cheng,~\IEEEmembership{Member,~IEEE}
       and Raymond W. Yeung,~\IEEEmembership{Fellow,~IEEE}
\thanks{F. Cheng  is with the Institute of Network Coding, The Chinese University of Hong Kong, N.T., Hong Kong. Email: fcheng@inc.cuhk.edu.hk}
\thanks{R. W. Yeung is with the Institute of Network Coding and the Department of Information Engineering, The Chinese University of Hong Kong, N.T., Hong Kong, and with the Key Laboratory of Network Coding Key Technology and Application and Shenzhen Research Institute, The Chinese University of Hong Kong, Shenzhen, China. Email: whyeung@ie.cuhk.edu.hk}
\thanks{This work was partially funded by a grant from the University Grants Committee of the Hong Kong Special Administrative Region (Project No.\ AoE/E-02/08) and Key Laboratory of Network Coding, Shenzhen, China (ZSDY20120619151314964). This paper was presented in part at Network Coding (NetCod), 2011.}
}

%


\maketitle

\begin{abstract}
Consider a communication network represented by a directed graph
$\mathcal{G}=(\mathcal{V},\mathcal{E})$, where $\mathcal{V}$ is the
set of nodes  and $\mathcal{E}$ is the set of point-to-point
channels in the network. On the network a secure message $M$ is
transmitted, and there may exist wiretappers who want to obtain
information about the message. In secure network coding, we aim to
find a network code which can protect the message against the
wiretapper whose power is constrained. Cai and Yeung
\cite{cai2002secure} studied the model in which  the wiretapper can
access any one but not more than one set of channels, called a
wiretap set, out of a collection $\mathcal{A}$ of all possible
wiretap sets. In order to protect the message, the message needs to
be mixed with a random key $K$. They proved tight fundamental
performance bounds when $\mathcal{A}$ consists of all subsets of
$\mathcal{E}$ of a fixed size $r$. However, beyond this special case,
obtaining such bounds is much more difficult. In this paper, we investigate
the problem  when $\mathcal{A}$ consists of arbitrary subsets of
$\mathcal{E}$ and obtain  the following results: 1) an upper bound
on $H(M)$; 2) a lower bound on $H(K)$ in terms of $H(M)$. The upper
bound on $H(M)$ is explicit, while the lower bound on $H(K)$ can be
computed in polynomial time when $|\mathcal{A}|$ is fixed. The tightness of the lower bound for
the point-to-point communication system is also proved.
\end{abstract}

\begin{IEEEkeywords}
Information inequality, perfect secrecy, performance bounds, secure network coding.
\end{IEEEkeywords}



\section{Introduction}
\IEEEPARstart{I}{n}
 classical information-theoretic cryptography, when we need to
send a private message to a receiver in the presence of wiretappers,
in order to protect the message, we  encrypt the message with a
random key and send the ciphertext to the receiver.\ A wiretapper
who has no access to the key can know nothing about the message by
only observing the ciphertext, in the sense that the ciphertext and
the message are statistically independent.\ On the other hand, the
receiver  obtains the key via a ``secure'' channel and use it to
decrypt the ciphertext to recover the private message. The best
known such model is the one-time pad system studied by Shannon
\cite{shannon1998communication}, which requires the minimal amount
of randomness for the key.

The one-time pad system was  generalized to \textit{secret sharing}
by Blakley \cite{blakley1979} and Shamir \cite{shamir1979share}.
Ozarow and Wyner \cite{ozarow1985wire} also studied a similar
problem which they called the \textit{wiretap channel~II}.  In this
model, information is sent to the receiver through a number of
point-to-point channels.  It is assumed that the wiretapper can
access any one but not more than one set of channels, called a
wiretap set, out of a collection $\cal A$ of all possible wiretap
sets, where $\cal A$ is specified by the problem under
consideration. For example, $\cal A$ could be the collection of all
wiretap sets  each containing a single channel. In this case, the
wiretapper can access any one but not more than one channel. The
strategy to protect the private message is the same as that in
classical information-theoretic cryptography. Specifically, the
private message and the random key are combined by means of a coding
scheme, so that a wiretapper observes some mixtures of the message
and the key, where these mixtures  are statistically independent of
the message. On the other hand, the receiver node can decode the
message from the information received on all the channels. Note that in secret sharing and its subsequent generalizations, it is assumed that the key is available only to the transmitter and transmission in all the channels is noiseless.

Cai and Yeung \cite{cai2002secure} generalized secret sharing to
secure network coding, in which a private message is sent  to
possibly more than one receiver through a network of point-to-point
channels. The model they studied, which we refer to as the
\textit{wiretap network} 
(see also El Rouayheb and Soljanin
\cite{rouayheb2007wiretap}), is described as follows. In this model,
the assumptions about the wiretapper and the strategy to protect the
private message are the same as in the wiretap channel II.
The significant difference is that there exist intermediate nodes in the
network that can encode, and there may be more than one receiver
node.  The solution is that we send both  the private message and
the key  via a network coding scheme, so that a wiretapper can only
observe some  mixtures of the message and the key, where the
mixtures  are statistically independent of the message. On the other
hand, a receiver node can recover the private message by decoding
the  information received from its input channels. Note that when
$\mathcal{A}$ is the empty set, the wiretap network reduces to the
original network coding model studied in Ahlswede \textit{et al.}
\cite{ahlswede2000network}.

In \cite{cai2002secure}, a condition for the existence of secure
linear network codes was proved and a construction of such codes was
proposed. The code in \cite{cai2002secure} suffers from the pitfall that the required alphabet size is larger than $|\mathcal{A}|$. Feldman \textit{et~al.} \cite{feldman2004capacity} generalize and simplify the method in \cite{cai2002secure}. They derived trade-off between security, the alphabet size and the multicast rate. Under their result, the alphabet size in \cite{cai2002secure} can be greatly reduced if a small amount of overall capacity is given up.
In  \cite{rouayheb2007wiretap}, El Rouayheb and Soljanin regarded the secure network coding problem as a network generalization of the model in wiretap channel II and showed that
the transmitted information can be secured by using the   coset coding scheme in \cite{ozarow1985wire}  at
the source on top of the existing network code. Moreover, their code is equivalent to the code in \cite{cai2002secure} but the required alphabet size is much smaller. The optimal code constructions in \cite{cai2002secure}, \cite{feldman2004capacity}, and  \cite{rouayheb2007wiretap} have a common strategy: they first construct a code on the message and the key at the source node and then transmit the source code via a network code, which  depends on the code at the source node.  In Cai \& Yeung
\cite{cai2007security} and Zhang \& Yeung \cite{zhang2009general},
a general security condition for multi-source  network code was
presented.

The performance of a secure network coding scheme is measured by two
quantities: the size of the message and the size of the key. In
designing a secure network coding scheme, we want to maximize the
size of the message and at the same time minimize the size of the
key. The latter is necessary because in cryptography, randomness is regarded
as a resource.  In \cite{cai2002secure},  when the
collection $\cal A$ of all wiretap sets consists of all subsets of
channels whose sizes are at most some constant $r$, an upper bound on the size of the message
and a lower bound on the size of the random key were obtained. Both of these bounds are tight for this special case. In this
paper, we extend these bounds to the general case.

Cui \textit{et al.}\ \cite{CuiHoWiretap} studied secure network coding in a single-source single-sink network with unequal channel capacities. The set of wiretap sets is arbitrary and randomness can be generated at the intermediate nodes. The aim is to find the maximal source-sink communication rate, i.e., the secrecy capacity. They give a cut-set bound on the  secrecy capacity and show that the cut-set bound is not achievable in general. Some achievable strategies are proposed and the computational complexity to determine the secrecy capacity is studied.

Secure network coding was also generalized from different perspectives. Bhattad and Narayanan \cite{BhattadNarayanan} introduced weakly secure network coding, where it is required that  wiretappers cannot decode any part of the source message. In this model, a weakly secure network code  can be used to avoid trading off the throughput. In \cite{Haradayamatao}, Harada and Yamamoto studied the  strongly $r$-secure linear network code which can protect the source message such that a wiretapper can obtain no information about any $s$ components of the source message by accessing $n-s$ channels provided that the maximum flows to all the sink nodes are at least $n$, where $s\leq n-r$. A polynomial-time algorithm was proposed to construct the strongly $r$-secure linear network code. They also showed that strong security contains  weak security as a special case.

Secure network coding with error correction was studied by Ngai and Yeung \cite{NgaiYeung09}, where they proposed a construction of secure error-correcting (SEC) network code which can protect the message from wiretapping, random errors and errors injected by the wiretapper. They further showed the optimality of their construction.

Security network coding was also well studied from a different point of view,  see \cite{lima2007random}--\cite{medard2006secure} for other related results.

%

\section{Problem Formulation}\label{sec:2}
In this work, we focus on the wiretap network model
proposed in \cite{cai2002secure} and aim to obtain some new  performance
bounds. Denote the network by $\mathcal{G}$ $=$ ($\mathcal{V}$, $\mathcal{E}$), where $\mathcal{V}$ is the set of nodes and
$\mathcal{E}$ is the set of edges, each representing a
point-to-point noiseless channel in the network. In this work, we use the
terms ``edge'' and ``channel'' interchangeably. On each edge $e$, a
symbol from some transmission alphabet $F$ can be transmitted. In
this sense we say that each channel has unit capacity. We assume
that $\cal G$ is a directed acyclic multigraph, namely there can be
multiple edges between each pair of nodes.

A wiretap network consists of the following components:
\begin{itemize}
  \item[1)] \textit{Source node $s$:} The node set $\mathcal{V}$ contains a
node $s$, called the source node, where a random message $M$ taking
values in an alphabet $\mathcal{M}$, called the message set, is
generated.

  \item[2)] \textit{Set of user nodes $\mathcal{U}$:} A user node is a node in
$\mathcal{V}$ which is fully accessed by a legal user who is
required to receive the random message $M$ with zero error. There is
generally more than one user node in a network. The set of user
nodes is denoted by $\mathcal{U}$. For each $u \in \mathcal{U}$, let
${\it maxflow(u)}$ denote the value of a maximum flow from the
source node s to node $u$.

  \item[3)] \textit{Collection of sets of wiretap edges $\mathcal{A}$:}
$\mathcal{A}$ is a collection of arbitrary subsets of the edge set
$\mathcal{E}$, called a wiretap set. The wiretapper can access any  $A\in \mathcal{A} $
but not more than one subset in $\mathcal{A} $ at the same time.
The wiretap set $A$ chosen by the wiretapper is fixed before communication. The sender can know $\mathcal{A}$ before communication but cannot figure out the exact $A$.
\end{itemize}
We denote such a wiretap network by the tuple  $(\mathcal{G}$, $s$, $\mathcal{U}$, $\mathcal{A})$.
\subsection{Admissible Code}
 We assume that the
message $M$ is generated at the source node according to an
arbitrary distribution on the message set $\mathcal{M}$. Let $K$ be
a random variable independent  of $M$, called the \textit{key}, that
takes values in an alphabet $\mathcal{K}$ according to the uniform
distribution.

For each node $v$ of the  network $\cal G$, we denote the set of the
input edges and the set of the output edges of $v$ by $\textit
In(v)$ and $\textit Out(v)$, respectively. A code for a wiretap
network consists of a set of local encoding mappings $\{\phi_e:e\in
\cal E \}$ such that for all $e$, $\phi_e$ is a function from $\cal
M \times \cal K$ to $F$ if $e \in \textit Out(s)$, and is a function
from $F^{|In(t)|}$ to $F$ if $e\in \it Out(t)$ for $t\neq s$. For
$e\in \cal E$, let $Y_e$ be the random symbol in $F$ transmitted on
channel $e$; i.e., the value of $\phi_e$. For a subset $B$ of $\cal
E$, denote $(Y_e:e\in B)$ by $Y_B$.

To complete the description of a code, we have to specify the order
in which the channels send the indices, called the \textit{encoding
order.} Since the graph $\cal G$ is acyclic, it defines a partial
order on the node set $\cal V$. Then the nodes in $\cal V$ can be
indexed in a way such that for two nodes $t$ and $t'$, if there is a
channel from node $t$ to node $t'$, then $t < t'$. According to this
indexing, node $t$ sends indices in its output channels before node
$t'$ if and only if $t < t'$. The order in which the channels within
the set of output channels of a node send the indices is immaterial.
The important point here is that whenever a channel sends an index,
all the indices necessary for encoding have already been received. A
code defined as such induces a function $\Phi_u$ from $\cal M \times
\cal K$ to $F^{|\it In(u)|}$ for all user nodes $u\in \cal U$, where
the value of $\Phi_u$ denotes the indices received by the user node
$u$ in its input channels.

In the wiretap network model, a code $\{ \phi_e: e \in \cal E  \}$
should satisfy the following two conditions:
\begin{itemize}
    \item[1)] \textit{decodable condition:} For all user node $u\in \cal
    U$ and all $m, m'\in \cal M$ with $m\neq m'$,
    \begin{center}
        $\Phi_u(m,k)\neq \Phi_u(m',k')$
    \end{center}
    for all $k, k'\in \cal K$. This guarantees that any two message
    are distinguishable at every user node.
    \item[2)]\textit{secure condition:} the message should be information-theoretic secure, namely
    for all
 $A\in \mathcal{A}$,
\begin{equation}\label{securecond}
H(M|Y_A) =H(M).
\end{equation}
We would like to emphasize that the wiretappers can know the encoding and decoding functions of the message and the key at all the nodes.
\end{itemize}
 We refer to a code satisfying 1) and 2) as an \textit{admissible code}.

For an admissible code, we focus on the following two performance
parameters, the size of the message and the size of the key:
\begin{itemize}
\item[1)] the size of the message is measured by $H(M)$, which should be maximized;
\item[2)] the size of the key is measured by $H(K)$, which should be minimized.
\end{itemize}
Furthermore, we can define an achievable region for $H(M)$ and $H(K)$, and what we have done in this paper is to characterize this region.
\subsection{Related Works} For set $A\subseteq B$, if $|A|=r$, then
we refer to it as an $r$-$subset$ of $B$.  In \cite{cai2002secure},
the following result was obtained.

\begin{Theorem}\label{re1}
    Let $q$ be the size of  the transmission alphabet $F$, $\mathcal{A}$ consist of all the r-subsets of
    $\mathcal{E}$ and $n =\min\limits_{u\in \mathcal{U}} \rm{maxflow}(u)$. Then
\begin{itemize}
\item [1)]    $H(M)\leq (n-r)\log q$;
\item [2)]  $H(K)\geq \frac{r}{n-r} H(M)$.
\end{itemize}
Moreover, when $F$ is a finite field such that $q>|\mathcal{A}|$, there exists a linear
admissible code which can achieve equalities in  these two bounds
simultaneously; i.e., the size of the message is maximized and the
size of the key is minimized.
\end{Theorem}
\begin{figure}
\begin{center}
\includegraphics[height= 180pt, width = 300pt]{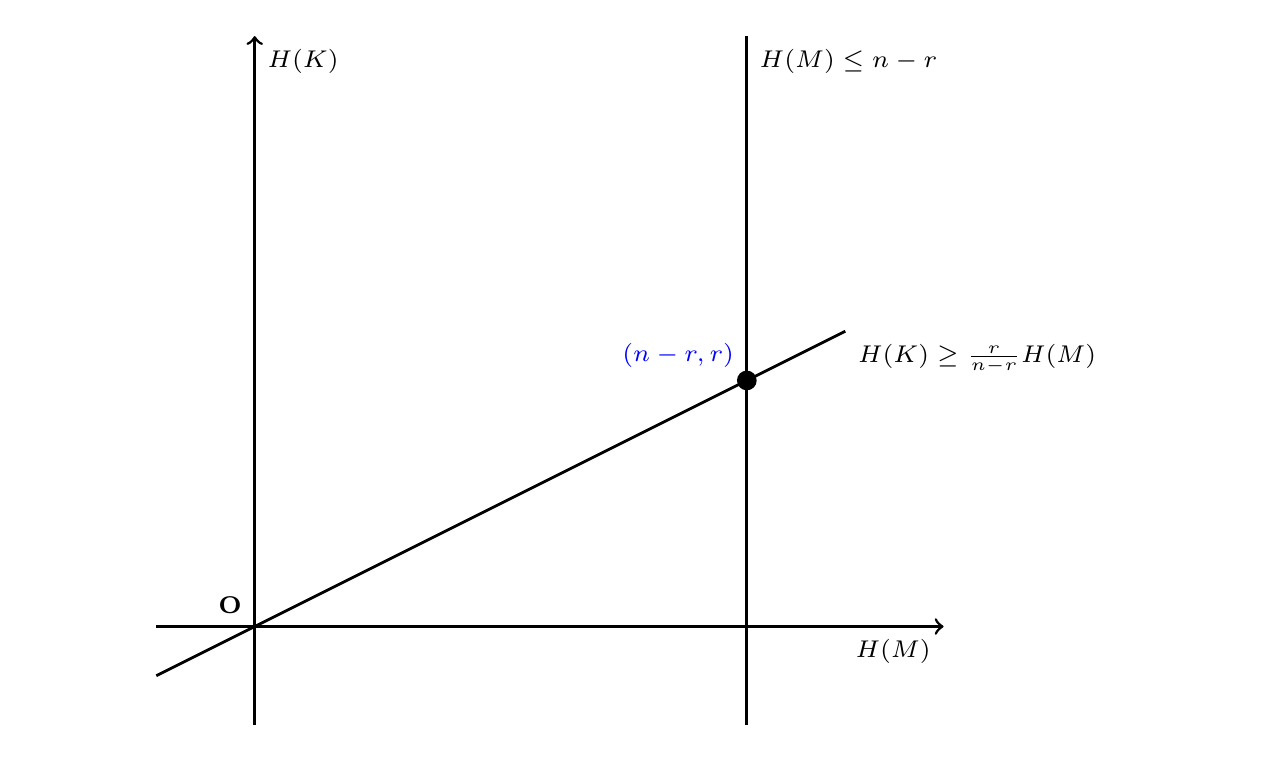}
\caption{The achievable region of $(H(M),H(K))$.}\label{mk}
\end{center}
\end{figure}

If all the logarithms are in the base $q$, then $1)$ becomes
\begin{itemize}
\item [$1')$] $H(M)\leq n-r.$
\end{itemize}
Fig.\ref{mk} illustrates the region of all ($H(M)$, $H(K)$) that satisfy $1')$ and $2)$. If time-sharing is allowed, then this is also the region of all achievable ($H(M)$, $H(K)$) because $(n-r$, $r)$ can be achieved by the code constructed in Cai \& Yeung.

However, when $\mathcal{A}$ consists of arbitrary subsets of
$\mathcal{E}$, the problem becomes very hard and very little is
known about the fundamental performance limit.
\begin{Example}
In Fig.\ref{jpg-butterfly}, the source node is $S$ and there are two destination nodes $U_1$ and $U_2$, the channel set is $\mathcal{E} = $$\{$$e_1$, $e_2$, ...,  $e_9$$\}$, where the channel capacity is unit. The message $M$ and the key $K$ are generated at $S$, and then are sent through the channels to $U_1$ and $U_2$. $M$ is required to be decodable at both $U_1$ and $U_2$.
If the set of wiretap sets $\mathcal{A}$ is $\{W: W\subseteq \mathcal{E}, |W|=1\}$, and for the wiretapper, it can access at most one of the sets in $\mathcal{A}$,
then the optimal sizes of the message and the key are known in $\cite{cai2002secure}$.
If the set of wiretap sets $\mathcal{A}$ is arbitrary, e.g., $\mathcal{A}=\{\{e_1\},\{e_3\}, \{e_5, e_6\}, \{e_5, e_7\}, \{e_6,e_7\}\}$, then the bounds on $H(K)$ and $H(M)$ are unknown in the literature.
\begin{figure}
  \centering
  \includegraphics[width = 160pt]{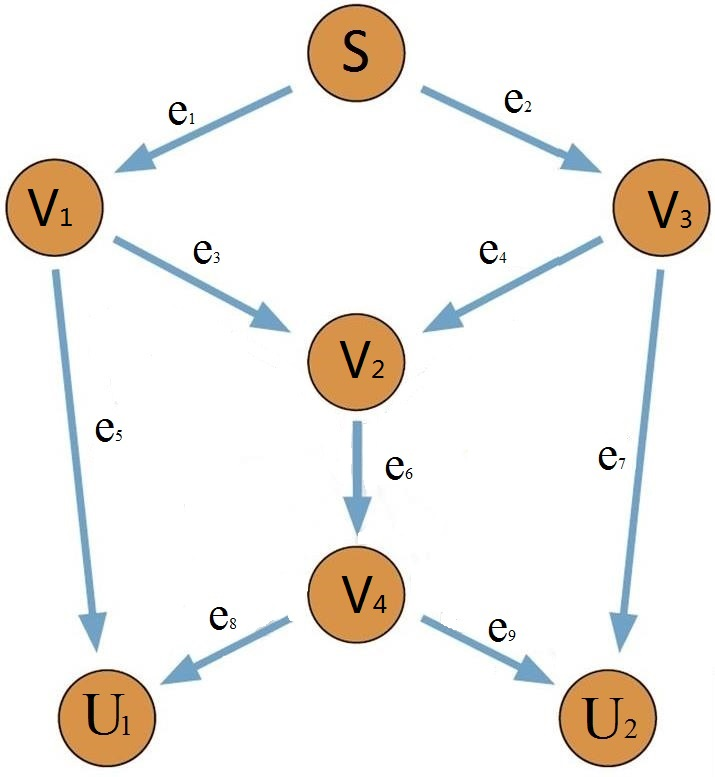}\\
  \caption{Secure network coding on the butterfly network.}\label{jpg-butterfly}
\end{figure}

\end{Example}

\subsection{Main Results}
In this work, we investigate the performance bounds when $\mathcal{A}$ is arbitrary. The main results are summarized as follows:
\begin{itemize}
  \item [1)] We obtain an upper bound on $H(M)$.
  \item [2)] We propose  a method to
  compute a lower bound on $H(K)/H(M)$, namely we obtain a lower bound  on $H(K)$
in terms of $H(M)$. We first propose a brute-force algorithm for
computing the lower bound. Then by refining the brute-force
algorithm, we obtain an algorithm whose computational complexity is
polynomial in $|\cal V|$ and $|\cal E|$ when  $|\mathcal{A}|$ is fixed. The lower bound
obtained by these algorithms is generally not tight. Nevertheless,
we prove that is tight for the classical point-to-point
communication system.
\end{itemize}

In the following  sections,  we first prove an upper bound  on
$H(M)$ in Sections \ref{sec:3} -- \ref{sec:4} and a lower bound on $H(K)$ in Sections \ref{sec:5} -- \ref{def1}. Then we discuss the algorithms
to compute these bounds in Section \ref{sec:10}. In Section \ref{sec:11}, we discuss the tightness of the
lower bound on $H(K)$. At last, we conclude the paper in Section \ref{sec:12}.

\section{Blocking Sets and Wiretap Sets}\label{sec:3}
In this section, we introduce some notations and theorems, which will be used to prove our results later.
\begin{defi}
For a network $\cal G=(\cal V,\cal E)$, we denote a cut (graph cut) of $\cal G$
by $(W, W^c)$, where $W\subseteq V$ contains the source node $s$ and $W^c=V\setminus W$
contains the destination node $t$, and denote the set of edges from
$W$ to $W^c$ by $E(W, W^c)$, which is also abbreviated to $E_W$.
\end{defi}

We first state in the next lemma two key inequalities obtained in
\cite{cai2002secure}.

\begin{lemma} \label{cyineq}
In the network $\cal G=(\cal V, \cal E)$, let $(W, W^c)$ be a cut of
$\cal G$. $\mathcal{A}$ consists of all the $r$-$subsets$ of $\mathcal{E}$. If there exists an admissible code on $\cal G$, then for
any wiretap set $I\subseteq E_W$, we have
\begin{itemize}
\item [$(A_1)$] $H(M)\leq H(Y_{E_W\setminus I} | Y_I)$;
\item [$(A_2)$] $H(K) \geq H(Y_I).$
\end{itemize}
\end{lemma}
The inequality $(A_1)$ was used in \cite{cai2002secure} to prove 1)
and 2) of Theorem~\ref{re1}. The inequality $(A_2)$ was  proved but
no further interpretation was provided.  In this section, we extend
these two inequalities to a more general situation where $\mathcal{A}$ is arbitrary. In the following discussion, unless otherwise stated, $I$ is assumed to be a wiretap set in $\mathcal{A}$.

\begin{defi}
In the network $\cal G$ $=$ $(\cal V, \cal E)$, a set $J$ $\subseteq
\cal E$ is called a blocking set if and only if there exists a cut
$(W, W^c)$ such that $E(W, W^c)\subseteq J$.
\end{defi}

The blocking set is a generalization of the graph cut. Let $u\in
\mathcal{U}$. Since the message $M$ can be decoded at user node $u$
and the symbols received at node $u$ are functions of $Y_{E_W}$,
where $W$ is a cut and $E_W$ is a subset of the blocking set $J$, we obtain that $M$
is a function of $Y_J$, namely
\begin{equation}
    H(M|Y_J) = 0.
\end{equation}

\begin{prop}\label{sec1}
Let $A, B\subset \cal E$ such that $B\subset A$. If $H(M|Y_A)=H(M)$, then
$H(M|Y_B)=H(M).$
\end{prop}
\begin{IEEEproof}
If $H(M|Y_A) = H(M)$, and $B\subseteq A$, then $$H(M|Y_B)\geq
H(M|Y_A) = H(M).$$ On the other hand, $$H(M|Y_B)\leq H(M).$$ Hence
$$H(M|Y_B) = H(M),$$ which completes the proof.
\end{IEEEproof}

The next lemma is a simple generalization of Lemma \ref{cyineq},
which we will see is a very useful tool for obtaining performance
bounds for a general secure network coding problem.
\begin{lemma}\label{main}
In the network $\cal G=(\cal V, \cal E)$, let $J \subseteq \cal E$
be a blocking set. For any admissible code on $\cal G$ and   any
wiretap set $I\subseteq J$, we have
\begin{itemize}
  \item [$(B_1)$] $H(M)\leq H(Y_{J\setminus I} | Y_I)$;
  \item [$(B_2)$] $H(K) \geq H(Y_I).$
\end{itemize}
\end{lemma}

\begin{IEEEproof}
Since $J$ is a blocking set, we obtain that
\begin{equation} \label{fc1}
 H(M|Y_J)= 0.
\end{equation}
Since $I\subseteq J$ is a wiretap set and the code is secure, we
have
\begin{equation} \label{fc2}
H(M|Y_I) = H(M).
\end{equation}
It follows that

\begin{center}
$\begin{array}{ccl}
  H(M) & = & H(M|Y_I)-H(M|Y_J) \\
   & =     & I(M;Y_{J\setminus I}|Y_I) \\
   & \leq  & H(Y_{J\setminus I}|Y_I),
  \end{array}
$
\end{center}
which completes the proof of $(B_1)$.

Since $H(Y_I|M,K)=0$, $I\subseteq J$,  
and  $H(Y_I)=H(Y_I|M)$, we obtain
that
\begin{center}
$\begin{array}{ccl}
  H(Y_I) & =    & H(Y_I|M)-H(Y_I|M,K) \\
         & =    & I(Y_I;K|M)\\
         & \leq & H(K|M) \\
         & =    & H(K),
\end{array}$
\end{center}
which completes the proof of $(B_2)$.
\end{IEEEproof}

In the following, we will first prove an upper bound on $H(M)$. Then we will prove the lower bound on $H(K)$.

\section{An Upper Bound on  the Message Size}\label{sec:4}
From Lemma \ref{main}, we can immediately obtain an upper bound on
$H(M)$.
\begin{corol}\label{col1}
Let the size of the transmission alphabet $F$ be $q$. Let $J$ be a
blocking set and $I\subseteq J$ be a wiretap set. For any admissible
code on $\cal G $,
\begin{equation}\label{b1}
H(M)  \leq  \min\limits_{J,I:I\subseteq J} |J\setminus I|\log q.
\end{equation}
\end{corol}

\begin{IEEEproof}
By $(B_1)$ of Lemma \ref{main}, we have
\begin{align}
 H(M)  & \leq   H(Y_{J\setminus I}|Y_I) \nonumber \\
       & \leq  H(Y_{J\setminus I}) \notag \\
       & \leq   |J\setminus I|\log q.
\end{align}

Then the corollary is proved by minimizing over all $J,I$ such that
$I\subseteq J$,
$$
H(M)  \leq  \min\limits_{J,I:I\subseteq J} |J\setminus I|\log q.
\nonumber
$$
\end{IEEEproof}

From this bound, we see that if \textit{$J\setminus I=\emptyset$},
then the upper bound above vanishes, which implies  $H(M)=0$. This means that if there exists a wiretap set $I$  that contains a cut as its subset,
then the network cannot send any message, because $J$  can be taken
to be $I$ so that $|J \setminus I| = 0$.

Next we present two theorems for computing the upper bound on
$H(M)$.
\begin{lemma} \label{m-cut}
For any fixed wiretap set $I$,
\begin{equation}\label{eq164}
\min\limits_{J:I\subseteq J} |J\setminus I|=\rm{mincut}(\mathcal{
E}\setminus I),
\end{equation}
where $\rm{mincut}(\mathcal{ E} \setminus I)$ is the minimum cut of
graph $(\mathcal{ V}, \mathcal{E}\setminus I)$.
\end{lemma}
\begin{IEEEproof}
Let $(W,W^c)$ be a graph cut and  $E_W$ be the edges across the cut.
Then $J_W=E_W\bigcup I$ is a blocking set. If we consider only such
blocking sets $J_W$ for $J$ in (\ref{eq164}), we have
\begin{align}
 & \min\limits_{J,I:I\subseteq J} |J\setminus I| \leq  \min\limits_{J_W}| J_W\setminus I|\notag\\
=& \min\limits_{E_W}| E_W\setminus I|=\rm{mincut}(\mathcal{E}\setminus I). \label{cut1}
\end{align}
The last equation is due to the fact that $E_W\setminus I$
corresponds to the set of edges across a cut of
$\mathcal{E}\setminus I$, and vice versa.

Conversely, let $J_0$ be a blocking set including $I$ that minimizes
$|J\setminus I|$, and $E_W \subseteq J_0$. Then
\begin{align}
&\min\limits_{J,I:I\subseteq J} |J\setminus I| = |J_0\setminus I|
\geq |E_W\setminus I|\notag\\
\geq & \min\limits_{E_W} |E_W\setminus I|=
\rm{mincut}(\mathcal{E}\setminus I).\label{cut2}
\end{align}
Together with (\ref{cut1}), we can conclude the proof.
\end{IEEEproof}

From Lemma \ref{m-cut}, we  obtain the following corollary.
\begin{corol}\label{col2}
$$\min\limits_{J,I:I\subseteq J} |J\setminus I|= \min\limits_{I} \rm{mincut}(\mathcal{ E}\setminus I).$$
\end{corol}
By means of this corollary, since the \rm{mincut} of a graph can be
computed in $O(|\cal V|\cdot |\cal E|)$ number of steps, we can
compute the upper bound on $H(M)$ in Corollary \ref{col1} in $O(
|I|\cdot|\cal V|\cdot |\cal E|)$ number of steps.

\section{Information Inequalities for Joint Entropy}\label{sec:5}
In this section, we state and explain some information inequalities
that are instrumental in proving the lower bound on $H(K)$.

Let $[n] = \{1, 2, ..., n \}$. For a subset $\alpha \subseteq [n]$,
denote $(X_i, i\in \alpha)$ by $X_{\alpha}$. Let $\bar{\alpha} =
[n]\backslash \alpha$. In information theory, the following
independence bound for joint entropy (e.g., p. 29 in
\cite{yeung2008information}) is well known.
\begin{center}
$H(X_{[n]})= H(X_1, \dots, X_n ) \leq \sum\limits_{i=1}^{n} H(X_i)$.
\end{center}
This inequality  provides an upper bound on the joint entropy
$H(X_{[n]})$ in terms of the entropies of the individual random
variables. It is tight when the random variables $X_1, \dots, X_n$ are
mutually independent.

\subsection{Han's Inequalities}
Han \cite{han1978nonnegative} generalized the independence bound to
two sequences of inequalities, which are stated in the next two theorems.

\begin{Theorem}
For $k=1$, $2$, $\dots$, $n$, let
\begin{center}
$H_k= \frac{1}{\binom{n}{k}}\sum\limits_{\alpha : |\alpha| =
k}\frac{H(X_{\alpha})}{k}$.
\end{center}
Then
\begin{equation}
H_n \leq H_{n-1} \leq \dots \leq H_{1}.
\end{equation}
\end{Theorem}
In this theorem, $$H_n = \frac{1}{n} H(X_{[n]}) \leq H_1 =
\frac{1}{n} \sum\limits_{i=1}^{n} H(X_i)$$ is equivalent to the
independence bound. This sequence of inequalities was used in
\cite{SymmeMulitlevelDeversity99} to prove a converse coding
theorem in multilevel diversity coding.

\begin{Theorem}
For $k=1$, $2$, $\dots$, $n$, let
\begin{center}
$H_k'= \frac{1}{\binom{n}{k}}\sum\limits_{\alpha : |\alpha| =
k}\frac{H(X_{\alpha} | X_{ \bar{\alpha}})}{k}$.
\end{center}
Then
\begin{equation}
H_1'\leq H_2'  \leq \dots \leq H_n' =\frac{ H(X_{[n]}) }{n}.
\end{equation}
\end{Theorem}
This sequence of inequalities was used in proving  $2)$ in Theorem \ref{re1}.

\subsection{Madiman-Tetali's Inequalities}
In Han's inequalities, the term $H_k$ $(H_k')$ only involves the
joint entropy (conditional joint entropy) of the $k$-$subsets$ of
$X_{[n]}$. These inequalities have recently been generalized by
Madiman and Tetali \cite{madiman2008information}. In the following,
let $C$ be  an arbitrary collection of subsets of $[n]$.
\begin{defi}
A function $\alpha$: $C\rightarrow R^{+}$ is called a
\textit{fractional covering} if $\sum_{s\in C:i\in s}\alpha(s)\geq 1$ for each $i\in [n]$.
\end{defi}

\begin{defi}
A function $\beta : C\rightarrow R^{+}$ is called a
\textit{fractional packing}, if $\sum_{s\in C:i\in s}\beta(s)\leq 1$ for each $i\in[n]$.
\end{defi}

\begin{Theorem}
For any collection $C$ of subsets of  $[n]$, any fractional covering $\alpha$ and any fractional packing $\beta$,
\begin{equation}\label{eq1}
\sum\limits_{s\in C}\beta(s)H(X_{s}|X_{s^c})\leq H(X_{[n]})\leq
\sum\limits_{s\in C}\alpha(s)H(X_s).
\end{equation}
\end{Theorem}
In the rest of this work, we refer to the left hand side of the
inequality as the fractional packing inequality and the right hand
side of the inequality as the fractional covering inequality.

\begin{Example}
Let $n = 3$ and $C=\{C_1, C_2, C_3\}$, where $C_1=\{1, 2\}$,
$C_2=\{2, 3\}$ and $C_3=\{1, 3\}$.

By  Han's inequalities, we obtain that
\begin{align} \label{example1}
\frac{1}{2}H(X_{{1,2}}|X_{3}) + \frac{1}{2}H(X_{{2,3}}|X_{1}) +  \frac{1}{2}H(X_{{3,1}}|X_{2})  & \leq H(X_{{1,2,3}})  &  \nonumber \\
\leq
\frac{1}{2}H(X_{1,2})+\frac{1}{2}H(X_{2,3})+\frac{1}{2}H(X_{3,1}). & &
\end{align}
Let $\alpha_i=\alpha(C_i)$ and $\beta_i=\beta(C_i)$, $i = 1, 2, 3$.
By Madiman-Tetali's inequalities, we obtain that
\begin{align} \label{example2}
 &\beta_1 H(X_{{1,2}}|X_{3}) + \beta_2 H(X_{{2,3}}|X_{1}) +  \beta_3 H(X_{{3,1}}|X_{2})\notag \\
 &\leq\  H(X_{{1,2,3}})  \notag \\
 &\leq\   \alpha_1 H(X_{1,2})+ \alpha_2 H(X_{2,3})+ \alpha_3 H(X_{3,1})
\end{align}
holds for any  fractional covering $\alpha$ and  any fractional
packing $\beta$, namely
\begin{center}
$\alpha_1, \alpha_2, \alpha_3 \geq 0$, $\alpha_1+\alpha_3\geq 1$, $\alpha_2+\alpha_3\geq 1$, $\alpha_3+\alpha_1\geq 1$;\\
$\beta_1, \beta_2, \beta_3 \geq 0$, $\beta_1+\beta_3\leq 1$, $\beta_2+\beta_3\leq 1$, $\beta_3+\beta_1\leq 1$.
\end{center}
In particular, when $\alpha_i = \frac{1}{2} $ and $\beta_i =
\frac{1}{2} $ for all $i=1, 2, 3,$ (\ref{example2}) becomes
(\ref{example1}). This shows that Madiman-Tetali's inequalities are
more general than Han's inequalities.

When $C_1=\{1, 2\}$, $C_2=\{2, 3\}$, $C_3=\{2\}$, Han's inequalities
are not applicable,  while by Madiman-Tetali's inequalities, we have
\begin{align}
&\beta_1 H(X_{{1,2}}|X_{{3}}) + \beta_2 H( X_{{2,3}} | X_{{1}})
+\beta_3 H(X_{{2}} |X_{{1,3}})   \notag \\
&\leq\   H(X_{{1,2,3}})    \nonumber \\
&\leq\  \alpha_1 H(X_{{1,2}}) +\alpha_2 H(X_{{2,3}}) +\alpha_3
H(X_{{2}}), &{ } &
\end{align}
where
\begin{center}
$\alpha_1\geq 1$, $\alpha_1+\alpha_2 +\alpha_3\geq 1$, $\alpha_2\geq
1$, and $\alpha_1, \alpha_2, \alpha_3 \geq 0$;\\
$\beta_1 \leq 1$, $ \beta_1+\beta_2+\beta_3\leq 1$, $\beta_2 \leq
1$, and $\beta_1, \beta_2, \beta_3\geq 0.$
\end{center}
Recently, Jiang \textit{et al.} \cite{JiangML} have applied these inequalities to multilevel diversity coding.
\end{Example}

\section{The Fractional Packing Bound}\label{fracBound}
In this section, we prove a lower bound on $H(K)$ by means of the
fractional packing inequality in  $($\ref{eq1}$)$.
\begin{Theorem}\label{fbound}
Fix a blocking set $J$ and let $\beta$ be a fractional packing of \
$\{ J\setminus I: I\subseteq J, I\in \mathcal{A}\}$, then
\begin{equation}\label{lp1}
H(K)\geq   \max\limits_{\beta}\left(\sum\limits_{I\subseteq
J}\beta(J\setminus  I)-1\right)H(M)
\end{equation}
\end{Theorem}

\begin{IEEEproof}
By $(B_1)$ of Lemma \ref{main}, we have
\begin{equation}
 H(M)\leq H(Y_{J\setminus I}|Y_I).
\end{equation}
By inequality $($\ref{eq1}$)$, we obtain
\begin{center} $\sum\limits_{I\subseteq J}\beta(J\setminus
I)H(M)\leq \sum\limits_{I\subseteq J}\beta(J\setminus
I)H(Y_{J\setminus I}|Y_{I})\leq H(Y_J)$.
\end{center}
Hence,
\begin{equation}\label{ieq1}
H(Y_J)\geq \sum\limits_{I\subseteq J}\beta(J\setminus I)H(M).
\end{equation}
From the definition of an admissible code, no keys are generated and used at the intermediate nodes. Hence $Y_J$ is a function of $M$ and $K$.
Then,
\begin{eqnarray}
  H(M)+H(K) & \geq & H(M,K) \nonumber \\
            & =     & H(M,K,Y_J) \nonumber \\
            & \geq  & H(Y_J) \nonumber \\
            & \geq  & \sum\limits_{I\subseteq J}\beta(J\setminus
            I)H(M).
\end{eqnarray}
This implies,
\begin{equation}\label{l1}
H(K)\geq   \left(\sum\limits_{I\subseteq J}\beta(J\setminus
I)-1\right)H(M).
\end{equation}
Since $($\ref{l1}$)$ holds for any fractional packing $\beta$, we have
\begin{equation}
H(K)\geq \left(\max\limits_{\beta}\sum\limits_{I\subseteq
J}\beta(J\setminus  I)-1\right)H(M),
\end{equation}
which completes the proof.
\end{IEEEproof}

In order to evaluate the lower bound on $H(K)$, we need to consider
the following LP (linear program),
\begin{eqnarray}
 \max &  \sum\limits_{I\subseteq J}\beta(J\setminus I)  \label{lp0}   \\
  s.t.  & \sum_{I\subseteq J:i\in J\setminus  I}\beta(J\setminus I)\leq 1, \forall
  i\in J
  \nonumber.
\end{eqnarray}
In the following discussion, we define
$\tau(J)=\max\limits_{\beta}\sum\limits_{I\subseteq
J}\beta(J\setminus I)-1$ for a fixed blocking set $J$, and let
$\tau=\max\limits_{J}\tau(J)$. Since in (\ref{lp0}), any $\{\beta(J\setminus I):I\subseteq J\}$ satisfying
$$\beta(J\setminus I)\geq 0, \sum\limits_{I:I\subseteq J}\beta(J\setminus I)=1$$ is a feasible solution, we 
obtain that $\tau(J)\geq 0$ and $\tau\geq 0$.

\begin{corol}\label{col3}
$\tau(J)>0$ if and only if for each edge $e\in J$, $e$ is covered by
some wiretap sets.
\end{corol}

\begin{IEEEproof}
If $e\in J$ is not covered by any wiretap set, then for all wiretap
set $I$, $e\in J\setminus I$.  By the LP in $($\ref{lp0}$)$, we
obtain that the constraint from edge $e$  is
$\sum\limits_{i=1}^{d}\beta_i \leq 1$, where $d$ is the number of
wiretap sets. This constraint dominates any other constraint, and
the maximum is attained when this bound is tight. Hence, $\tau (J) =
\sum\limits_{i=1}^{d} \beta_{i}-1 = 0$.

Conversely, assume that for all $e\in J$, it is covered by at least one wiretap
set. Fix $e$, and we can assume that, without lost of generality,
$e\in I_1$. Then we have $e \notin J\setminus I_1$, implying that
the number of sets $J\setminus I_j\ (j\neq 1)$ which cover $e$ is at
most $d-1$. Let $\beta_i = \frac{1}{d-1}$ for $1\leq i \leq d$. Then
$\beta_i$ is a feasible solution, and hence
$\tau(J)\geq\sum\limits_{i=1}^{d} \beta_{i}-1=1/(d-1)>0.$
\end{IEEEproof}

Corollary \ref{col3} has the following implication. For a fixed $J$, if there exists an edge $e\in J$ such that $e$ is not covered by any wiretap set, then $\tau(J)=0$, and so
$$\tau = \max_{J'}\tau(J') = \max_{J'\neq J}\tau(J').$$
On the other hand, if every edge $e\in J$ is covered by at least one wiretap set, then $\tau(J)>0$, and so
$$\tau = \max_{J'}\tau(J') \geq \tau(J)>0.$$
Therefore, for the purpose of computing $\tau$, we assume without loss of generality that every edge $e\in J$ is covered by at least one wiretap set.

\section{An Alternative Bound}\label{sec:7}

In the last section, we proved a lower bound on $H(K)$ in terms of
fractional packings of $\{J\setminus I: I\subseteq J, I\in \mathcal{A} \}$ for all
blocking sets $J$. In this section, we prove an alternative lower
bound on $H(K)$ in terms of fractional coverings of $\{I: I\subseteq
J, I\in \mathcal{A}\}$. In the next section, we prove a duality result between
fractional packing and fractional covering that implies the
equivalence of  these two bounds.

Fix a blocking set $J$. By $(B_2)$ of Lemma \ref{main}, for any
wiretap set $I\subseteq J$, we have
\begin{equation}
H(K)\geq H(Y_I).
\end{equation}
Let $\alpha$ be a fractional covering of $\{I: I\subseteq J\}$. By
the fractional covering inequality in $($\ref{eq1}$)$, we obtain
that
\begin{equation}
H(Y_J)\leq \sum\limits_{I\subseteq J}\alpha(I)H(Y_I)\leq
\sum\limits_{I\subseteq J}\alpha(I)H(K).
\end{equation}

Together with $($\ref{ieq1}$)$, we further obtain
\begin{equation} \label{lower1}
\sum\limits_{I\subseteq J}\beta(J\setminus I)H(M) \leq H(Y_J)\leq
\sum\limits_{I\subseteq J}\alpha(I)H(K).
\end{equation}
Then
\begin{equation}
H(K) \geq \frac{\sum\limits_{I\subseteq J}\beta(J\setminus I)}{\sum\limits_{I\subseteq J}\alpha(I)} H(M).
\end{equation}
Maximizing over all $\beta$ and minimizing over all $\alpha$, we
obtain another lower bound on $H(K)$ for a fixed $J$:
\begin{equation}\label{lp2}
H(K)\geq \frac{\max\limits_{\beta}\sum\limits_{I\subseteq
J}\beta(J\setminus I)}{\min\limits_{\alpha}\sum\limits_{I\subseteq
J}\alpha(I)} H(M).
\end{equation}
The maximization in the above has been considered in Section~\ref{fracBound}. Thus in order to evaluate the above lower bound on
$H(K)$, we also need to consider the following LP:
\begin{eqnarray}\label{eqf}
 \min &  \sum\alpha(I)  \label{lp3}   \\
  s.t  & \sum_{I\subseteq J:i\in I}\alpha(I)\geq 1, \forall i \in J
  \nonumber.
\end{eqnarray}

\section{A Duality Result}\label{sec:8}
In this section, we prove that $($\ref{lp2}$)$ is equivalent to
$($\ref{lp1}$)$.

\begin{Theorem} \label{duality}
For a given  blocking set $J$,
$$\max\limits_{\beta}\left(\sum\limits_{I\subseteq J}\beta(J\setminus I)-1\right)
= \frac{\max\limits_{\beta}\sum\limits_{I\subseteq
J}\beta(J\setminus I)}{\min\limits_{\alpha}\sum\limits_{I\subseteq
J}\alpha(I)}, $$ where $\alpha$ is a fractional covering of $\{I:
I\subseteq J, I\in \mathcal{A}\}$ and $\beta$ is a fractional packing of  $\{
J\setminus I: I\subseteq J, I\in \mathcal{A}\}$.
\end{Theorem}
In the following discussions, let
\begin{equation}\label{def:lcj}
l_{C}(J)=\min\limits_{\alpha}\sum\limits_{I\subseteq J}\alpha(I)
\end{equation}
and
\begin{equation}\label{def:lpj}
l_{P}(J)=\max\limits_{\beta} \sum\limits_{\it{ I\subseteq}
J}\beta(J\setminus I),
\end{equation}
where $\alpha$ is a fractional covering of
$\{I: I\subseteq J\}$ and $\beta$ is a fractional packing of $\{
J\setminus I: I\subseteq J\}$.

\begin{IEEEproof} [Proof (Theorem \ref{duality})]
In this proof, since $J$ is fixed, we can use $l_C$ and $l_P$
instead of $l_C(J)$ and $l_P(J)$ without ambiguity. We need
to prove
\begin{equation}
l_{P}-1 = \frac{l_{P}}{l_{C}},
\end{equation}
namely
\begin{equation}\label{eq0}
 l_{C} = \frac{l_{P}}{l_{P}-1} \  \ or \ \  l_{P} =
\frac{l_{C}}{l_{C}-1} .
\end{equation}
Let $I_1$, $I_2$, $\dots$, $I_d$ be the wiretap sets in $J$.

\begin{enumerate}

\item [(1)] Let $\alpha^* =$\argmin$\left\{ {\sum\limits_{I\subseteq J}\alpha(I)}
\right\}$ and for each wiretap set $I_i$, $\alpha^*_i=\alpha^*(I_i)$.
For $1\leq i \leq d$, define
\begin{equation} \label{sum-beta}
sum =\sum\limits_{i=1}^{d}\alpha^*_i \text{ and } \beta^*_i=\frac{\alpha^*_i}{sum-1}.
\end{equation}
Next, we  prove that $\{\beta^*_i: 1\leq i \leq d\}$ is a feasible
solution to the  LP in $($\ref{lp0}$)$; i.e., there exists a fractional packing $\beta^*$ on $\{J\setminus I_i: 1\leq i\leq d\}$ such that $\beta^*(J \setminus I_i) =
\beta_i^*$.

For each $e\in J$, we can assume without loss of generality that  $I_1,$ $\dots$, $I_j$
are the sets containing $e$ and $I_{j+1}$, $\dots$, $I_d$ be the sets
not containing $e$. Since $\{\alpha^*_i: 1\leq i \leq d\}$ is a
fractional covering,  $\sum\limits_{i=1}^{j}\alpha^*(I_i)\geq 1$.
For every $e\in J$, since $e\notin J\setminus I_s$, for $1\leq s\leq j$
and
$e\in J\setminus I_s$, for $j+1\leq s \leq d$,  
 we have
\begin{align}
\sum\limits_{i=j+1}^{d}\beta^*(J\setminus I_i) &
=\sum\limits_{i=j+1}^{d}\frac{\alpha^*(I_i)}{sum-1}=\frac{\sum\limits_{i=j+1}^{d}\alpha^*(I_i)}{sum-1}\notag \\
&=\frac{sum-\sum\limits_{i=1}^{j}\alpha^*(I_i)}{sum-1}  \leq
\frac{sum-1}{sum-1}=1. \notag
\end{align}

Since  $l_{P}$ is the maximum of the summation in (\ref{def:lpj}) over all fractional packing $\beta$,  together with (\ref{sum-beta}), we have
\begin{equation}\label{d1}
    l_{P} \geq \sum\limits_{i=1}^{d}\beta^*_i = \frac{sum}{sum-1} =
    \frac{l_{C}}{l_{C}-1}.
\end{equation}

\item[(2)] Let $\beta^* =$\argmax$\left\{\sum\limits_{I\subseteq
J}\beta(J\setminus I)\right\}$ and  for each wiretap set $I_i$, $\beta^*_i=\beta^*(J\setminus I_i)$.
For $1\leq i\leq d$, define
\begin{equation}\label{sum-alpha}
sum = \sum\limits_{i=1}^{d}\beta^*_i \text{ and } \alpha_i^*=\frac{\beta^*_i}{sum-1}.
\end{equation}
Next, we  prove that $\{\alpha^*_i : 1\leq i \leq d\}$ is a feasible
solution to the  LP in (\ref{lp3}); i.e., there exists a fractional covering $\alpha^*$ on $\{I_i: 1 \leq i \leq d\}$ such that $\alpha^*(I_i) =\alpha^*_i$.

For each $e\in J$, we can assume without loss of generality that $I_1$, $\dots$, $I_j$ are
the sets containing $e$ and $I_{j+1}$, $\dots$, $I_d$ be the sets not
containing $e$. Since $\{\beta_i^*: 1\leq i\leq d\}$ is a fractional
packing,  $\sum\limits_{i=j+1}^{d}\beta^*(J\setminus I_i)\leq 1$.
For every $e\in J$, since $e\notin J\setminus I_s$, for $1\leq s\leq j$
and $e\in J\setminus I_s$, for $j+1\leq s \leq d$, we have
\begin{align}
\sum\limits_{i=1}^{j}\alpha^*(I_i)&=\sum\limits_{i=1}^{j}\frac{\beta^*(J\setminus
I_i)}{sum-1}=\frac{\sum\limits_{i=1}^{j}\beta^*(J\setminus
I_i)}{sum-1} \notag\\
&=\frac{sum-\sum\limits_{i=j+1}^{d}\beta^*(J\setminus I_i)}{sum-1}
\geq \frac{sum-1}{sum-1}=1.\notag
\end{align}
Since $l_{C}$ is the minimum of the summation in (\ref{def:lcj}) over all fractional covering $\alpha$,  together with (\ref{sum-alpha}), we have
\begin{equation}\label{d2}
    l_{C} \leq \sum\limits_{i=1}^{d}\alpha^*_i = \frac{sum}{sum-1} =
    \frac{l_{P}}{l_{P}-1}.
\end{equation}
By $($\ref{d1}$)$ and $($\ref{d2}$)$, we obtain $l_{C}l_{P}\geq
l_{C}+l_{P}\geq l_{C}l_{P}$, namely $l_{C}l_{P} = l_{C}+l_{P}$,
which completes the proof.
\end{enumerate}
\end{IEEEproof}

By Theorem \ref{fbound} and \ref{duality}, the following   bound is equivalent to the bound in Theorem \ref{fbound}.

\begin{Theorem}\label{fcbound}
Fix a blocking set $J$ and let $\alpha$ be a fractional covering of $\{I: I\subseteq J, I\in \mathcal{A}\}$. Then
\begin{equation}\label{coveringbound}
H(K)\geq \max\limits_{\alpha}\frac{1}{\sum\limits_{I\subseteq
J}\alpha(  I)-1} H(M).
\end{equation}
\end{Theorem}

By $($\ref{eq0}$)$, we can  write the lower bound in Theorem
\ref{fbound} or \ref{fcbound} as $\frac{H(K)}{H(M)}\geq \frac{1}{l_C
-1}$ and consider only the LP in (\ref{eqf}). Since $\tau =
\max\limits_{J}1/(l_{C}(J)-1)=\max\limits_{J}(l_{P}(J)-1)$, we need
to find $\min\limits_{J}l_{C}(J)$ or $\max\limits_{J} l_{P}(J)$. In
the following sections, we refer to these two equivalent bounds as the
\textit{fractional covering bound} and the \textit{fractional packing bound},
respectively.

\section{Some Properties of the Lower Bound}\label{def1}

Consider the matrix form of the LP in $($\ref{lp3}$)$ for the
fractional covering. Let $I_1$, $I_2$, $\dots$, $I_d$ be the wiretap sets. For
each blocking set $J$,  construct a $|J|\times d$ matrix $A_J$ to
represent the edges in $J$  as follows. Let $e_1^J$, $e_2^J$, $\dots$, $e_{|J|}^J$ be the edges
in $J$. If $e_i^J\in I_j$, then $A_J(i,j) = 1$, else $A_J(i,j)=0$.
Each column of $A_J$ corresponds to a wiretap set, and each row of
$A_J$ corresponds to an edge in $J$.

We can now write the LP in $($\ref{lp3}$)$ and its dual as
\begin{center}
$\begin{array}{ccccccccc}
 {\rm LP}: & \min &  & 1^Tx    &  &{\rm Dual}: & \max & & 1^Ty\\
      & s.t & & A_Jx\geq \textbf{1} &  &     &  s.t  & & A_J^Ty\leq \textbf{1}\\
      &  & & x\geq 0     &  &     &     & & y\geq 0
\end{array}
$
\end{center}
The strong duality theorem in linear programming (Theorem \ref{a4}
in the appendix) states that the LP and its dual
problem have the same optimal value.

When we try to solve the above LP, we need to consider some special
relations among the wiretap sets and the blocking sets, namely  a
wiretap set is a subset of another wiretap set, or a blocking set is
a subset of another blocking set. We discuss these issues in the following.

\begin{corol}
For a given blocking set $J$,  if wiretap sets $I_i$ and $I_j$
satisfy $I_i\subseteq I_j\subseteq J$, then $I_i$ can be ignored in
the model.
\end{corol}
\begin{IEEEproof}
For wiretap sets $I_i$, $I_j\subseteq J$, if $I_i \subseteq I_j$,
then the \textit{i}th and \textit{j}th column of $A_J$ satisfy
$A_J^i\leq A_J^j$ componentwise, which implies in the dual problem
the constraint $(A_J^i)^Ty\leq 1$ is dominated by the constraint
$(A_J^j)^Ty\leq 1$. Thus we can  ignore the column $A_J^i$ in $A_J$,
or equivalently, the wiretap set $I_i$.
\end{IEEEproof}
In the following discussion, we assume that $I_i$ $(1\leq i \leq d)$
is not a subset of any other wiretap sets.

\begin{corol}
If the blocking sets $J'$, $J$ satisfy $J'\subseteq J$, then
$\tau(J) \leq \tau(J').$
\end{corol}
\begin{IEEEproof}
By definition, if $J' \subseteq J$, then $A_{J'}$ is a submatrix of
$A_J$. By comparing the linear programs for $J'$ and $J$, we notice
that the two objective functions  are the same, but the feasible
region of $J$ is a subset of that of $J'$, because $A_{J'}$ is a
submatrix of $A_J$. Since we need to obtain the minimum value of the
objective function, we have   $l_{C}(J')\leq l_{C}(J)$, where
$l_{C}(J')$ and $l_{C}(J)$ are the optimal values for $J'$ and $J$,
respectively. Then  $\tau(J')=1/(l(J')_{C}-1)\geq
1/(l(J)_{C}-1)=\tau(J)$, which concludes the proof.
\end{IEEEproof}
This corollary implies that toward computing $\tau=\max\limits_{J}\tau(J)$,
if $J'\subseteq J''$, then $J''$ can
be ignored. In particular, since each blocking set contains a graph
cut (also a blocking set), toward computing $\tau$, it is attained over all graph cuts between the source and destination
nodes.

In the following sections, we will discuss the algorithms on computing the bound on $H(K)/ H(m)$ and the tightness of our bound.

\section{Algorithms for Computing the Lower Bound}\label{sec:10}

\subsection{A Brute Force Algorithm}
Based on the above discussion, we  propose a brute force algorithm,
namely that  we enumerate all the graph cuts and solve the
corresponding LPs (e.g., by the simplex algorithm). Then the time
complexity is $2^{|\mathcal{V}|}O(LP)$, where $O(LP)$ is the time
complexity of the LP; e.g., the interior point algorithm can terminate in $O(m^2n+m^3)$ arithmetic operations, where $m$ is number of constraints and $n$ is the number of variables.
\begin{Theorem}\label{Sperner}
Sperner's Theorem \cite{van2001course}: If $A_1, A_2, ..., A_m$ are
subsets of $N:=\{1,2,...,n\}$ such that $A_i$ is not a subset of
$A_j$ if $i\neq j$, then $m\leq
\binom{n}{\lfloor\frac{n}{2}\rfloor}$.
\end{Theorem}

When solving the LP, the primary factors of the complexity are the
number of variables and constraints, namely the number of  wiretap
sets $d$ and $|J|$. By Theorem \ref{Sperner}, since for every two
wiretap sets $I_i$ and $I_j$, $I_i$ is not a subset of $I_j$ if
$i\neq j$, we obtain  $d\leq
\binom{|J|}{\lfloor\frac{|J|}{2}\rfloor} $.

In this algorithm, the total complexity is exponential, which is not
practical when the problem size is large. Next we  propose an
algorithm which is polynomial when $d$ is  constant.

\subsection{A Polynomial-Time Algorithm }\label{palgo}

In this part we show that when the number of wiretap sets, $d$, is a
constant, there exists a polynomial algorithm for computing the
lower bound. In the following discussion, we use some definitions
and theorems in  linear optimization which are given in the appendix.

In the above brute force algorithm, we consider every blocking set
$J$ and solve the following linear program for $J$:
\begin{center}
$\begin{array}{cccl}
  {\rm LP}(J): & \min &  & 1^Tx     \\
      & s.t. &  & A_Jx\geq 1   \\
      &     &  & x\geq 0,x\in R^d.
\end{array}
$
\end{center}
If we let $A_J'=\left(
                  \begin{array}{c}
                    A_J \\
                    I_{d\times d} \\
                  \end{array}
                \right)
$ and $b_J=\left(
      \begin{array}{c}
        1_{|J|} \\
        0_d \\
      \end{array}
    \right)
$, where $I_{d\times d}$ is the $d\times d$ identity matrix, then
the above constraints can be written as $A_J'x\geq b_J$.

Let $P=\{x\in R^d \ | \ A_Jx\geq 1,x \geq 0\}$. Since $x=1_d\in P$,
$P$ is a nonempty polyhedron. Since $A_J'$ contains $I_{d\times d}$
as a submatrix, we see that there exist $d$ rows of $A_J'$ which
are linearly independent. So by Theorem \ref{a2} (in Appendix \ref{App1}),  the
polyhedron $P$ has at least one extreme point. Since $x\geq 0$,  the
optimal value is nonnegative, and hence not equal to $-\infty$. By
Theorem \ref{a3}, there exists an extreme point which is optimal.
Let $x^*(J)$ denote an extreme point (not necessary unique) that
gives the optimal solution. Then by Theorem \ref{a1}, $x^*(J)$ is a
basic feasible solution. A straightforward method to find $x^*(J)$
is to enumerate all the basic solutions of $LP(J)$, and check
whether the basic solutions are feasible or not. In order to
enumerate all the basic feasible solutions, we consider all $d\times
d$ submatrices of $A_J'$. For such a submatrix $S$, there is a
corresponding basic solution if and only if $\rank(S)=d$, and if so,
denote this basic solution by $x_S$. Then a basic solution $x_S$ is
feasible if $A_J'x_S\geq b_J$. Among all these basic feasible
solutions, $x^*(J)$ is the one that attains the minimum value.

To sum up, we draw the following conclusion.

\begin{con}
For blocking set $J$, the optimal solution can be obtained by
solving one of the equations: $Sx=b_S$, where $S$ is a $d\times d$
submatrix of $A_J'$ and $b_S$ is the corresponding subvector of
$b_J$.
\end{con}

Furthermore, to obtain the best lower bound on $\frac{H(K)}{H(M)}$,
we need to solve the linear program to obtain the optimal value for
each blocking set. Then take the minimum of these optimal values
over all blocking sets to obtain the best lower bound. This can be
achieved by repeating the procedure in Conclusion~$1$.

The method described above is inefficient because if $S$ is a
submatrix  of both $A_{J_1}'$ and $A_{J_2}'$ for two different
blocking sets $J_1$ and $J_2$, the exact  same processing of $S$
would be performed twice. In the remaining of this section, we aim
to improve the method by removing such redundant operations.

In the method described above, if we obtain the best lower bound on
$\frac{H(K)}{H(M)}$ from blocking set $J$, we refer to the optimal
value and the optimal solution of  $LP(J)$ as the best optimal value
and the best optimal solution. Recall that for each blocking set
$J$, since $J\subseteq \mathcal{E}$, $A_J'$ is a submatrix of
$A_{\mathcal{E}}'$ ($\mathcal{E}$ is a blocking set so
$A_{\mathcal{E}}'$ is defined accordingly). Then we can draw another
conclusion.

\begin{con}\label{conclusion2}
Consider the best lower bound on $\frac{H(K)}{H(M)}$ in network
$\mathcal{G}=(\mathcal{V},\mathcal{E})$. The best optimal solution
can be obtained by solving one of the equations $Sx=b_S$, where $S$
is a $d\times d$ submatrix  of $A_{\mathcal{E}}'$ and $b_S$ is the
corresponding subvector of $b_{\mathcal{E}}$.
\end{con}
\begin{defi}\label{def:Q}
For each blocking set $J$, let $Q_J$ be the set of all  basic
feasible solutions of $LP(J)$, and let $Q =\bigcup\limits_{J} Q_J $.
\end{defi}

Let $\gamma=\binom{|\mathcal{E}|+d}{d}$. By Conclusion
\ref{conclusion2}, the best optimal value is $\min\limits_{x\in
Q}1^Tx$. If we compute the set $Q$ by means of the prescription in
Definition \ref{def:Q}, we need to enumerate all the blocking sets,
and hence the computational complexity is exponential in
$|\mathcal{E}|$. But we notice that matrix $A_{\mathcal{E}}'$ has
$\gamma$ submatrices with dimension $d\times d$ and each of them
corresponds to at most one basic feasible solution, and so $|Q|\leq
\gamma$. When $d$ is a constant, $\gamma$ is polynomial in
$|\mathcal{E}|$, which suggests that if we compute $Q$ by
enumerating these $\gamma$ $d\times d$ submatrices,  we may obtain
an algorithm which is polynomial in $|\mathcal{E}|$. By the
definition of $Q$, for each $d\times d$ submatrix $S$, if
$\rank(S)<d$, we cannot obtain a basic solution from $Sx=b_S$.
Therefore, we only need to consider $S$ such that
\begin{itemize}
\item[1)] $\rank(S)=d.$
\end{itemize}
When $S$ satisfies $1)$, $Sx=b_S$ has a unique solution, which we
denote  by $x_S$. In the sequel, whenever we discuss $x_S$, we
implicitly assume that $S$ satisfies $1)$, otherwise $x_S$ is
undefined. If $x_S$ is feasible for some blocking set $J$, namely
$A_J'x_S\geq b_J$, then $x_S$  satisfies
\begin{itemize}
\item[2)] $x_S\geq 0$.
\end{itemize}
Let $Q'$ be the set of all $x_S$ satisfying $2)$. Then $Q\subseteq
Q'$ and $Q'$ can be computed in polynomial time. Now we need to
solve the following problem: if $x\in Q'$, what is the necessary and
sufficient condition for $x\in Q$?



For each edge $e\in \mathcal{E}$, let $(a^e)^T$ denote the row of
$A_{\mathcal{E}}$ corresponding to $e$. For each $x_S\in Q'$, let
$F(S)=\{e\in \mathcal{E} |(a^e)^Tx_S\geq 1\}$.

\begin{Theorem}\label{theorem9}
Let $x_S\in Q'$. Then $x_S\in Q$  if and only if $F(S)$ is a
blocking set.
\end{Theorem}

\begin{IEEEproof}
``$\Rightarrow$'' For $x_S\in Q'$, if $x_S\in Q$, then $x_S$ is a
basic feasible solution of  $LP(J)$ for some blocking set $J$. By
$A_J'x_S\geq b_J$, we obtain that for each $e\in J$, $(a^e)^Tx_S\geq
1$, which means $e\in F(S)$, implying $J\subseteq F(S)$. Hence
$F(S)$ is a blocking set.

``$\Leftarrow$''  Recall that $A_{\mathcal{E}}'=\left(
                  \begin{array}{c}
                    A_J \\
                    I_{d\times d} \\
                  \end{array}
               \right )$. For a $d\times d$ submatrix $S$ of $A_{\mathcal{E}}'$,
let $E_S$ be the set consisting of all $e\in \mathcal{E}$ such that
$e$ corresponds to a row of $S$. By the definition of $x_S$, we have
that for each $e\in E_S$, $(a^e)^Tx_S=1$, which means that $e\in
F(S)$, implying that $E_S\subseteq F(S)$. Let $J=F(S)$. Then $J$ is
a blocking set. For $e\in J$, $(a^e)^Tx_S\geq 1$, namely $A_Jx_S\geq
1$. Together with $x_S\geq 0$, we have $A_J'x_S\geq b_J$. Since
$Sx_S=b_S$ and $S$ is a $d\times d$ submatrix of $A_J'$, $x_S$ is a
basic feasible solution of $LP(J)$, and hence $x_S\in Q$.
\end{IEEEproof}

By Theorem \ref{theorem9}, for $x_S\in Q'$, in order to determine
whether $x_S\in Q$, we only need to  check whether $F(S)$ is a
blocking set. This can be done in polynomial time as follows. In the
graph $\mathcal{G}=(\mathcal{V},\mathcal{E})$, upon deleting all the
edges in $F(S)$,  we  need to check whether the source node and the
destination node are connected  in the residual graph, which can be
achieved via a Depth-First Search (DFS) algorithm (e.g., in
 \cite{cormen2001introduction}) with time
complexity $O(|\mathcal{V}|+|\mathcal{E}|)$.
Based on the these results, we propose Algorithm 1 on the next page
for computing the lower bound on $H(K)/H(M)$.
\begin{algorithm}\label{algorithm:1}
\caption{Algorithm for computing a lower bound on
$\frac{H(K)}{H(M)}$}
\begin{itemize}
  \item [a)] For each $d\times d$ submatrix $S$ of $A_{\mathcal{E}}'$, keep the
 matrix provided that it satisfies  $\rank(S)=d$ and $x_S\geq
0$.
  \item [b)] For each  $S$ that survives in a), calculate $F(S)$, and
  determine whether  $F(S)$ is a blocking set. If
so, calculate $\rm val(S)= 1_d^Tx_S$, else ignore $S$.
   \item [c)] Output $S$ and $x_S$ that attain the  minimum $\rm val(S)$.
\end{itemize}
\end{algorithm}

The time complexity analysis of Algorithm 1 is as follows:
\begin{itemize}
\item [1.] In step $a)$, the time for calculating all $x_S$ is $O(\gamma*d^3)$,
where $d ^3$ is the time for matrix inversion by Gaussian
elimination.

\item [2.] In step $b)$, in the worst case, we need to enumerate all the
$\gamma$ submatrices. For each submatrix $S$, there are at most
$|\mathcal{E}|$ edges in $F(S)$, and so we have to delete at most
$|\mathcal{E}|$ edges in graph
$\mathcal{G}=(\mathcal{V},\mathcal{E})$. The complexity for
determining whether $F(S)$ is a blocking set  is
$O(|\mathcal{V}|+|\mathcal{E}|)$. In sum, the time complexity of
this step is $O(\gamma*(|\mathcal{V}|+|\mathcal{E}|))$.

\item [3.] With steps a) and b) together, the total complexity is
$O(\gamma*d^3+\gamma*(|\mathcal{V}|+|\mathcal{E}|))=
O(|\mathcal{E}|^{d}(|\mathcal{V}|+|\mathcal{E}|))$, which is
polynomial when $d$ is a constant.
\end{itemize}

\section{Tightness of the Lower Bound}\label{sec:11}
In this section, we discuss tightness of the lower bound on
$H(K)/H(M)$ obtained by  Algorithm $1$. In Cai and Yeung
\cite{cai2007security}, a security condition for multi-source linear
network coding was proved. This condition, stated in the next
theorem, is instrumental in the discussion in this section. For the
sake of completeness, we include in Appendix~\ref{App2} a proof of this theorem which is somewhat simpler than the proof in \cite{cai2007security}.

In the sequel, let $F_q$ be a finite field of size q and $F_q^r =
\underbrace{F_q\times F_q ... \times F_q}_{r}$. For a matrix $A$, we
also write the number of rows and columns of $A$ as $\row(A)$ and
$\col(A)$, respectively.

\begin{Theorem} \label{s1}
Let $A$ and $B$ be given matrices defined on $F$. Let $M$ be a random vector with positive probability distribution on $F_q^{m}$ and $K$ be a uniformly distributed random vector on $F_q^{k}$.
 Let
$Y=\left(\begin{array}{ll}
             A & B\\
             \end{array}\right )
          \left(
         \begin{array}{l}
           M \\
           K \\
         \end{array}
       \right)
$ and  $C=\left(
         \begin{array}{cc}
           A & B \\
         \end{array}
       \right)$ and assume that $\rank(C)$ is equal to the number of rows
of $C$.
 Then the following are equivalent:
\begin{itemize}
  \item [a)] $M$ and $Y$ are independent, namely $I(Y;M)=0$;
  \item [b)] $\rank(B)=\row(B)$, or equivalently, $\rank(B)=\rank(C)$.
\end{itemize}
\end{Theorem}
In practice, when $q\to\infty$, the matrix $C$ can be generated randomly. With high probability approaching $1$, $\rank(C)$ is equal to the rows of $C$.

\subsection{When the Best Lower bound is Zero}
In this case, the lower bound on $H(K)/H(M)$ is tight as we
now show. By $\tau=\max\limits_{J}\tau(J)=0$, we obtain that for
each blocking set $J$, $\tau(J)=0$. In Corollary \ref{col3}, by
letting $J$ be an arbitrary graph cut $(W,W^c)$ of network
$\mathcal{G=(V,E)}$, we see that there exists an edge $e\in
E(W,W^c)$ such that $e$ is not contained in any wiretap set. Hence
in $\mathcal{G=(V,E)}$, if we delete all the edges which are
contained in some wiretap sets, then the number of  remaining edges
in each graph cut is at least $1$. By the max-flow min-cut theorem,
there exists a path $P$ from the source node to the destination node
and all the edges in $P$ are not contained in any wiretap sets. So
we can send a message  $M$ along $P$ without mixing it with a random
key. For such a  scheme, $H(M)>0$ and $H(K)=0$, implying that the
bound $H(K)/H(M)\geq 0$ is tight.

\subsection{Point-to-Point Communication System}\label{sectionp2p}
In this section, we prove that in a point-to-point communication
system, the lower bound on $H(K)/H(M)$ is tight. Consider
such a system. Let $s$ and $u$ be the source node and the
destination node, respectively. Let $h$ be the number of edges from
node $s$ to node $u$ and  $I_1,$ $I_2$, $\dots,$ $I_d$ be the wiretap sets. 

We now write the $LP$ in (\ref{eqf}) and its dual as follows
\begin{align}\label{lp4}
  {\rm Primal}: & \min &  & 1^Tx     &  {\rm Dual}: & \max  & & 1^Ty\notag \\
      & s.t & & A_Jx\geq \mathbf{1}    & &  s.t & & A_J^Ty\leq \mathbf{1} \notag\\
      &  & & x\geq 0,x\in R^d & &      & & y\geq 0, y\in R^h
\end{align}
Since the primal  has an optimal solution $x^*$,  by the strong
duality theorem in linear optimization (Theorem \ref{a4} in the appendix), the dual also has an optimal solution
$y^*$ and $1^Tx^*=1^Ty^*$. Next we prove that the lower bound on
$\frac{H(K)}{H(M)}$ can be achieved, namely there exists a code such
that $H(M) = (1^Ty^*-1)H(K)$.
\begin{prop}
There exists an optimal solution $y^*$ such  that all its entries
are rational numbers.
\end{prop}
\begin{IEEEproof}
By Conclusion $1$, there exists an extreme point $y^*$ which is
optimal. This extreme point can be obtained by solving a particular set of  linear equations, whose coefficients are rational
numbers. Hence we conclude that $y^*$ is also rational.
\end{IEEEproof}
\bigskip
Let $y^*=(a_1/b_1,$  $a_2/b_2,$ $\dots$, $a_h/b_h)^T,$ where $a_i, b_i\in \mathds{N}$ and
$gcd(a_i,b_i)=1$, $1\leq i \leq h$. Let $g=lcm(b_1$, $b_2$, ...,
$b_h)$, and $w_i=g\cdot a_i/b_i$, $w_i\in \mathds{N}$. Let
$w_{\max}=\max\limits_{1\leq i \leq h}w_i$ and $w =
\left(\sum\limits_{i=1}^{h} w_i-g\right)$. Then
$1^Ty^*-1=\frac{w}{g}$. Let $M$ and $K$ be uniformly distributed on
$F_q^g$ and $F_q^w$, respectively. Next, we prove that there exists
a linear code with transmission alphabet $F=F_q^{w_{\max}}$  such
that $H(K)=g$ and $H(M)=w$ (where the logarithm is in the base $q$),
and on each edge $e_i$ $(1\leq i\leq h)$, the codeword is a vector
defined on $F_q^{w_i}$. By appending to the codeword a zero vector
of length $w_{\max}-w_i$, the codeword becomes a vector in $F$. When
$w_i= 0$, we transmit nothing on edge $e_i$, so we can ignore edge
$e_i$. In the following, without loss of generality, we assume that
$w_i>0$.

\begin{prop}\label{prop3}
There exists a wiretap set $I$ such that $$\sum\limits_{e_i\in I}w_i
= g.$$
\end{prop}
\begin{IEEEproof}
Since $y^*$ is a basic feasible solution of the dual problem in
(\ref{lp4}), we can find matrix C such that
\begin{equation}\label{eq100}
Cy^* =
    \left(
          \begin{array}{c}
            1_{n_1} \\
            0_{n_2} \\
          \end{array}
    \right),
\end{equation}
where $C$ is an invertible $h\times h$ submatrix of $\left(
          \begin{array}{c}
            A_J^T \\
            I_{h\times h} \\
          \end{array}
        \right)$  and $n_1+n_2=h$. In the dual problem, we can see
that $y_0=(1, 0, \dots, 0)\in R^h$ is a feasible solution and
        $1^Ty_0=1$. Therefore, $1^Ty^*\geq 1^Ty_0=1$. If $n_1=0$,
        then $y^*=0$, so that $1^Ty^*=0$, a contradiction. Hence,
        $n_1>0$.
Then we obtain from (\ref{eq100}) that
\begin{equation}
C\left(
    \begin{array}{c}
      w_1 \\
      w_2 \\
      \vdots \\
      w_h\\
    \end{array}
  \right)
=\left(
    \begin{array}{c}
      g \\
      g \\
      \vdots \\
      0\\
    \end{array}
  \right).
\end{equation}
Letting $I$ be the wiretap set that corresponds to the first row of
$C$, we have $\sum\limits_{e_i\in I}w_i = g$.
\end{IEEEproof}

Without loss of generality, we can let the wiretap set  $I$
prescribed in Proposition \ref{prop3} be $I_d=\{e_{t+1}$, $e_{t+2}$, $\dots$, $e_h\}$,
so that the edges apart from those in $I_d$ are $e_1$, $e_2$, $\dots$, $e_t$.
Then for each $I_i$ where $1\leq i\leq d-1$, by $A_J^Ty^*\leq 1$ and
$y^*=(w_1/g$, $w_2/g$, $\dots$, $w_h/g)$,  we have
\begin{equation}\label{w1}
 \sum\limits_{j:e_j\in I_i}w_j\leq g
\end{equation}
for  $1\leq i \leq d-1$.

We assume \begin{equation}M=\left(
                \begin{array}{c}
                  m_1 \\
                  m_2 \\
                  \vdots \\
                  m_t \\
                \end{array}
              \right)
\in F^{w}_q,
\end{equation} where $m_i\in F^{w_i}_q$ $(1\leq i \leq
t)$.
Let $B_i$ ($1\leq i\leq h$) be a $w_i\times g$ matrix defined on
$F_q$ to be specified later. Let the symbol  transmitted on edge
$e_i$ be
\begin{equation}\label{key1}
Y_i = m_i+B_iK,
\end{equation}
where  $Y_i\in F_q^{w_i}$,  $1\leq i \leq t$, and let
\begin{equation}
Y_{I_d} =B_{I_d}K,
\end{equation}
where
\begin{equation}
B_{I_d}=\left(
                                     \begin{array}{c}
                                       B_{t+1} \\
                                       B_{t+2} \\
                                       \vdots \\
                                       B_{h} \\
                                     \end{array}
                                   \right)
\end{equation}
is the $g\times g$ identity matrix on $F_q$. Namely, for $t+1\leq
i\leq h$, the symbol transmitted  on edge $e_i$ is
\begin{equation}
Y_i=B_iK.
\end{equation}
Let $Y$ be the symbols transmitted on all the edges. Then we can
write
\begin{align}
 Y & =\left(
      \begin{array}{c}
        Y_1 \\
        Y_2 \\
        \vdots  \\
        Y_t \\
        Y_{I_d} \\
      \end{array}
    \right)\notag\\
  &=\left(
                     \begin{array}{cccccc}
                       D_1 & 0   & 0  &...      & 0       &B_1 \\
                       0   & D_2 & 0  &...      & 0       &B_2 \\
                      \vdots  &\vdots   &\vdots  &\vdots       & \vdots      &\vdots  \\
                       0   & 0   &... & 0       & D_{t}   &B_{t}\\
                       0   & 0   & ...& 0       & 0       &B_{I_d} \\
                     \end{array}
                   \right)
\left(
   \begin{array}{c}
     m_1 \\
     m_2\\
     \vdots  \\
     m_t \\
     K\\
   \end{array}
 \right)\label{y}
\end{align}
where  $D_i$, $1\leq i \leq t$, is the $w_i\times w_i$ identity
matrix.


For a matrix $A$, we denote the vector space spanned by the rows of
$A$ by $\rowspan(A)$. For each $e_i$ ($1\leq i\leq h$), let
$V_i=\rowspan(0,...,D_i,...,$ $0, B_i)$ (the row space of the
\textit{i}th row in (\ref{y})). From the above construction, we have
$\dim(V_i)=w_i$ for $1\leq i\leq h$ and $\dim( V_1\oplus V_2\oplus
...$ $ \oplus V_h)=\sum\limits_{i=1}^{h}w_i$.

In the code we have constructed, we see from (\ref{key1}) that the
$g$ symbols of the key $K$ are sent on the edges in $I_d$.
Therefore, $I(Y_{I_d};M) = 0$. The following lemma, which is a refinement of  Lemma 3 in
\cite{cai2002secure}, is instrumental for constructing $B_i, 1\leq
i\leq t$.
\begin{lemma}\label{c1}
Let $V_1$, $V_2$, ..., $V_m$ be  vector subspaces on $F^n_q$, and
$\dim(V_i)$ $ =d_i$ $(1\leq i\leq m)$. If $d\geq 0$ and $d+d_i\leq
n$ $(1\leq i\leq m)$, then for  $q>m$, there exists a vector
subspace $V$ of $F^n_q$, such that $\dim(V)=d$ and $\dim(V \oplus
V_i) = \dim(V)+\dim(V_i)$ $(1\leq i\leq m)$.
\end{lemma}
\begin{IEEEproof}
Let $\{b_1,b_2, ..., b_d\}$ be a  basis of $V$. For all $1\leq i\leq m$, let  $\{v_{i1},v_{i2},..., v_{id_i}\}$ be a maximally
independent set of vectors in $V_i$. We construct $\{b_1,b_2, ..., b_d\}$ by induction. It suffices to show that for
$1\leq j\leq d$, if $b_1, b_2, ..., b_{j-1}$ have been chosen such
that for all $V_i$, $1\leq i\leq m$,
\begin{equation}\label{20120206:1}
b_1, b_2, ..., b_{j-1},v_{i1},v_{i2},..., v_{id_i}
\end{equation}
are linearly independent, then it is possible to choose $b_j$ such
that for all $1\leq i\leq m$,
\begin{equation}
b_1, b_2, ..., b_{j-1}, b_j, v_{i1},v_{i2},..., v_{id_i}
\end{equation}
are linearly independent. Specifically, $b_j$ is chosen such that it
is independent of the set of vectors in (\ref{20120206:1}) for all
$1\leq i\leq m$; i.e.,
\begin{equation}
    b_j\in F_q^n\setminus\cup_{1\leq i\leq m}\langle b_1, b_2, ..., b_{j-1},v_{i1},v_{i2},..., v_{id_i}\rangle.
\end{equation}
Since the cardinality of a subspace in $F_q^n$ is finite, we need to
show that the set above is nonempty.
\begin{align}
&\bigg|\bigcup\limits_{1\leq i\leq m}\langle b_1, b_2, ..., b_{j-1},v_{i1},v_{i2},..., v_{id_i}\rangle\bigg|\notag\\
\leq &\ \sum\limits_{1\leq i\leq m}\bigg|\langle b_1, b_2, ..., b_{j-1},v_{i1},v_{i2},..., v_{id_i}\rangle\bigg|\notag\\
=&\ \sum\limits_{1\leq i\leq m}q^{d_i+j-1} \notag\\
\leq &\ \sum\limits_{1\leq i\leq m}q^{n-1}\ ({\rm for}\ d_i+j\leq d_i+d\leq n)\notag\\
= &\ mq^{n-1}.\notag
\end{align}
Therefore,
\begin{align}
&\ \bigg|F_q^n\setminus\bigcup\limits_{1\leq i\leq m}\langle b_1, b_2, ..., b_{j-1},v_{i1},v_{i2},..., v_{id_i}\rangle\bigg|\notag \\
\geq &\ q^n -mq^{n-1}\notag \\
= &\ q^{n-1}(q-m)\notag \\
> &\ 0, \notag
\end{align}
since $q> m$. Hence $b_j$ can be chosen for all $1\leq j\leq m$.
\end{IEEEproof}

In the following, we  construct $B_i, 1\leq i \leq t$ to satisfy the
secure condition: for each wiretap set $I$, $I(Y_I; M)=0$. Since the
symbols transmitted on the edges in wiretap set $I_i= \{e_{i_1},
e_{i_2}, ..., e_{i_{|I_i|}}\} $ $(1\leq i\leq d-1)$ are
    \begin{align}&\left(
                                                \begin{array}{c}
                                                    Y_{i_1} \\
                                                    Y_{i_2} \\
                                                    \vdots  \\
                                                    Y_{i_{|I_i|}} \\
                                                \end{array}
                                                \right)=\\&\left(
                                                \begin{array}{ccccccc}
                                                    0& ...& D_{i_1} & ...   & ... &...  &  B_{i_1} \\
                                                    0& ...& ...      &D_{i_2}& ... & ... & B_{i_2} \\
                                                    \vdots &\vdots &\vdots &\vdots &\vdots &\vdots &\vdots  \\
                                                    0& ...& ... &... & D_{i_{|I_i|}}& ...&    B_{i_{|I_i|}} \\
                                                \end{array}
                                                \right)\left(
    \begin{array}{c}
        m_1 \\
        m_2\\
        \vdots  \\
        m_t \\
        K\\
    \end{array}
    \right),
    \end{align}
by Theorem \ref{s1}, for each wiretap set $I_i$ $(1\leq i\leq d-1)$,
if
    \begin{equation}
    T_i=\left(
    \begin{array}{c}
        B_{i_1} \\
        B_{i_2}\\
        \vdots  \\
        B_{i_{|I_i|}} \\
    \end{array}
    \right)
    \end{equation}
satisfies b) of Theorem \ref{s1}, namely
\begin{equation}\label{cond1}
\dim(T_i) = \row(T_i)=\sum\limits_{j=1}^{|I_i|}\row(B_{i_j})=
\sum\limits_{j=1}^{|I_i|} w_{i_j},
\end{equation}
then for $I_i$, the secure condition holds.

For $1\leq i \leq d-1$, we define matrix $T_i^0$ as follows: if
$I_i\cap I_d=\{e_{j_1}, e_{j_2}, ..., e_{j_r}\}$, then
    \begin{equation}T_i^0=\left(
            \begin{array}{c}
                B_{j_1} \\
                B_{j_2} \\
                \vdots  \\
                B_{j_r} \\
            \end{array}
            \right),
    \end{equation}
else $T_i^0$ is the empty matrix. For each $i$, $T_i^l$ for $1\leq
l\leq t$ are defined inductively as follows: if $e_l\in I_i$, then
$$T_i^{l}=\left(
                                                             \begin{array}{c}
                                                               T_i^{l-1} \\
                                                               B_l \\
                                                             \end{array}
                                                           \right)
,$$ else $$T_i^{l}= T_i^{l-1}.$$ We can verify that for $1\leq i\leq
d-1$, the rows of $T_i^{t}$ are a permutation of the rows of  $T_i$.
Hence, (\ref{cond1}) holds if and only if
    \begin{equation}\label{cond101}
    \dim(T_i^t) = \row(T_i^t).
    \end{equation}

Now, we construct $B_i, 1\leq i\leq t$ one by one starting from
$B_1$. For each $l$, $1\leq l\leq t$, we need to construct $B_l$
such that $T_i^l$ satisfies b) of Theorem \ref{s1}; i.e.,
    \begin{equation}\label{cond102}
    \dim(T_i^l) = \row(T_i^l),
    \end{equation}
for $1\leq i\leq d-1$.

Before we construct $B_1$, for wiretap set $I_i$ ($1\leq i\leq
d-1$), since $B_{I_d}$ is an identity matrix, if $I_i\cap I_d\neq
\emptyset$,   then
$$\dim(T_i^0) = \sum\limits_{j:e_j\in I_i\cap I_d}w_j=\row(T_i^0),$$
else $\dim(T_i^0) = 0$. For either case, (\ref{cond102}) holds.

For $B_1$,  $\row(B_1)=w_1$, and it is required that if $e_1\in
I_i$,
    \begin{align}
    \dim(T_i^1)&=\row(T_i^1)\notag\\
               &=\row(T_i^0)+\row(B_1)\notag\\
               &=\row(T_i^0)+w_1,\label{eq2}
    \end{align}
for $1\leq i\leq d-1$. By (\ref{w1}), if $e_1\in I_i$ ($1\leq i\leq
d-1$), we have
    \begin{align}
    \row(T_i^0)+w_1 &=   \sum\limits_{j:e_j\in I_i\cap I_d}w_j + w_1\notag\\
                    & \leq \sum\limits_{j:e_j\in I_i}w_j \notag \\
                    &  \leq g. \label{cond100}
    \end{align}
By (\ref{cond100}) and Lemma \ref{c1}, we can construct a $w_1
\times g$ matrix $B_1$ to satisfy (\ref{eq2}), and hence
(\ref{cond102}) is satisfied for $l=1$.

We assume that for a fixed $l'$, where  $1\leq l' \leq t-1$, $B_1,
B_2, ..., B_{l'}$ have been constructed so that (\ref{cond102}) is
satisfied for $1\leq l \leq l'$. Then
    \begin{equation}\label{cond4}
    \dim(T_i^{l'}) =\row(T_i^{l'})=\sum\limits_{j:e_j\in I_i\cap I_d}w_j+
    \sum\limits_{j:e_j\in I_i,j\leq l'}w_j.
    \end{equation}
For $B_{l'+1}$,  $\row(B_{l'+1})= w_{l'+1}$, and it is required that
if $e_{l'+1}\in I_i$,
    \begin{equation}\label{cond3}
    \dim(T_i^{l'+1})=\row(T_i^{l'+1})=\row(T_i^{l'})+w_{l'+1}.
    \end{equation}
By (\ref{w1}) and (\ref{cond4}), if $e_{l'+1}\in I_i$,
    \begin{align} \label{cond2}
    \row(T_i^{l'})+w_{l'+1}& = \sum\limits_{j:e_j\in I_i\cap I_d}w_j+ \sum\limits_{j:e_j\in I_i,\ j\leq l'+1}w_j\notag\\
    &\leq \sum\limits_{j:e_j\in I_i}w_j \notag \\
    &\leq g.
    \end{align}
By Lemma \ref{c1} and (\ref{cond2}), we can construct a
$w_{l'+1}\times g$ matrix $B_{l'+1}$ such that  (\ref{cond3}) holds,
and hence (\ref{cond102}) is satisfied for $l=l'+1$. By mathematical
induction, we can construct $B_i, 1\leq i \leq t$.

The decoding can be done as follows. We first obtain $K$ from
wiretap set $I_d$. Then $y_i$ can be solved for all $1\leq i\leq h$
and by (\ref{key1}) we obtain that $m_i=Y_i-B_iK$ for $1\leq i \leq
t$.

For the code we have constructed, $H(M)=w$ and $H(K)=g$, so that
$H(M)/H(K)=w/g = 1^Ty^*-1$ as desired. Hence the
lower bound on $H(K)/H(M)$ by Algorithm $1$ is tight.

Now, we give an example to demonstrate our lower bound on $H(K)/H(M)$.
\begin{Example}
Let $\{e_1,e_2,e_3\}$ be a cut-set. The set of wiretap sets $\mathcal{A}=\{A_1,A_2\}$, where $A_1=\{e_1,e_2\}, A_2=\{e_2,e_3\}$ are two wiretap sets.
By the fractional covering bound, we have
\begin{align}
    \max\ \ \ \ & x_1+x_2+x_3 \\
    s.t.\ \ \ \ & x_1+x_2\leq 1; \label{e2:1}\\
                & x_2+x_3\leq 1; \label{e2:2}\\
                & 0\leq x_1,x_2,x_3\leq 1;
\end{align}
It is easy to see $x_1=x_3=1, x_2=0$ is an optimal solution. Hence
\begin{equation}
\frac{H(K)}{H(M)}\geq \frac{1}{x_1+x_2+x_3-1}=1.
\end{equation}
Let $H(K)=1$. From our construction of the code that achieves the lower bound, we see that $x_i$ ($i=1,2,3$) can be interpreted as the information rate on channel $e_i$, with the information transmitted on channel $e_1$, $e_2$, and $e_3$ being mutually independent. The constraints  ($\ref{e2:1}$) and ($\ref{e2:2}$) mean the size of the symbols in each wiretap set cannot exceed the size of the key, which is similar to  Shannon's perfect secrecy.

On the other hand, we cannot directly apply the bounds in Cai \& Yeung \cite{cai2002secure} since $\mathcal{A}$ does not contain the set $\{e_1,e_3\}$. If we consider a weaker set of wiretap sets $\mathcal{A}' =\{A_1',A_2',A_3'\}$, where $A_{1}' = \{e_1\}, A_{2}' = \{e_2\},$ and $A_{3}' = \{e_3\}$. By the bounds in Cai \& Yeung \cite{cai2002secure}, we have
\begin{equation}
\frac{H(K)}{H(M)}\geq \frac{1}{2},
\end{equation}
which is strictly less than our bound.
\end{Example}

\section{Conclusion}\label{sec:12}
In this paper, we have obtained an upper bound on the size of the
message and  a lower bound on the size of the  key for a secure
network code on a wiretap network, when the set of wiretap sets $\mathcal{A}$ is arbitrary. The lower bound on the size of the key is obtained via a set of entropy inequalities by Madiman and
Tetali \cite{madiman2008information}. The bound on $H(K)$ consists of a fractional covering bound and a fractional packing bound, which can be proved to be equivalent. Computation of this bound can
be achieved in polynomial time when $|\mathcal{A}|$ is fixed, and it is tight for the special case
of the point-to-point communication system. That is, from the perspective of cut-set bound, our lower bound on $H(K)$ is optimal. Compared to the existing bounds, our bound is more general to outperform all of them. Consider the region of points $(H(M),H(K))$, our result has established an outer bound on the achievable region. Moreover,  our bounds have characterized the performance of routing, which is a special network code and can be simplified as a point-to-point communication system.

\ifCLASSOPTIONcaptionsoff
  \newpage
\fi

\appendices
\section{Linear Optimization}\label{App1}


In this appendix, we  present some  standard definitions and theorems in linear optimization  taken from \cite{bertsimas1997introduction}.

\begin{defi}
A polyhedron is a set that can be described in the
form $\{x\in R^n | Ax \geq b\}$, where A is an $m\times n$ matrix and b is a
vector in $R^m$.
\end{defi}

\begin{defi}
Let $P$ be a polyhedron. A vector $x\in P$ is an
extreme point of $P$ if we cannot find two vectors $y,z\in P$, both
different from $x$, and a scalar $\lambda \in [0,1]$, such that
$x=\lambda y+ (1-\lambda)z$.
\end{defi}

\begin{defi}
Let $P$ be a polyhedron. A vector $x\in P$ is a
vertex of $P$ if there exists some $c'$ such that $cx'<c'y$ for all
$y$ satisfying $y\in P$ and $y\neq x$.
\end{defi}

\begin{defi}
Consider a polyhedron $P$ defined by linear
equality and inequality constraints, and let $x^*$ be an element of
$R^n$.
\begin{itemize}
  \item [(a)] The vector $x^*$ is a basic solution if:
  \begin{enumerate}
    \item All equality constraints are active.
    \item Out of the constraints that are active at $x^*$, there are
    $n$ of them  that are linearly independent.
  \end{enumerate}
  \item [(b)] If $x^*$ is a basic solution that satisfies all of the
  constraints, we say that it is a basic feasible solution.
\end{itemize}
\end{defi}

\begin{Theorem}
\label{a1}
Let $P$ be a nonempty polyhedron and let $x^* \in P$. Then, the
following are equivalent:
\begin{itemize}
  \item [(a)] $x^*$ is a vertex;
  \item [(b)] $x^*$ is an extreme point;
  \item [(c)] $x^*$ is a basic feasible solution.
\end{itemize}
\end{Theorem}

\begin{defi}
A polyhedron $P \subset R^n$ contains a line if there exists a
vector $x \in P$ and a nonzero vector $d\in R^n$ such that
$x+\lambda d\in P$ for all scalars $\lambda$.
\end{defi}

\begin{Theorem}
\label{a2}
 Suppose that the polyhedron $P=\{x \in R^n |
a_{i}'x \geq b_i, i=1,...,m\}$ is nonempty. Then, the following
are equivalent:
\begin{itemize}
\item[(a)]  The polyhedron P has at least one extreme point.
\item[(b)]  The polyhedron P does not contain a line.
\item[(c)]  There exists n vectors out of the family $a_1,...,a_m,$
which are linearly independent.
\end{itemize}
\end{Theorem}

\begin{Theorem}
\label{a3}
 Consider the linear programming problem of minimizing
$c'x$ over a polyhedron P. Suppose that P has at least one extreme
point. Then, either the optimal cost is equal to $-\infty$, or there
exists an extreme point which is optimal.
\end{Theorem}

\begin{Theorem}[Strong duality]\label{a4}
If a linear programming problem has an optimal solution, so does its dual, and the respective optimal costs are equal.
\end{Theorem}


\section{Proof to Theorem \ref{s1}}\label{App2}
\begin{IEEEproof}
$``a)\Rightarrow b)"$ Since $\rank(C)=\row(C)$, we have
$$\row(C)\leq \col(C).$$
Then for each $Y=y$, the equation $y=AM+BK$
has at least one solution for $(M,K)$, which means
$$\Pr(Y=y)>0.$$
Together with $\Pr(M=m)>0$ and $I(Y;M)=0$, we obtain that
$$\Pr(Y=y,M=m)=\Pr(Y=y)\Pr(M=m)>0,$$
namely for each $y$ and $m$, the
equation $y=Am+BK$ has at least one solution for $K$. Since $BK=y-Am$ has at
least one solution for arbitrary $(y,m)$, we obtain
$\rank(B)=\row(B)$.

$``b)\Rightarrow a)"$ Let $W=AM$, $V=BK$ and $r=\rank(B)$. Since
$K$ is uniformly distributed, $V$ is uniformly distributed on
$F_q^r$. Since $\row(Y)=\row(V)$,
$$H(Y)\leq\log|F_q^r| = H(V)=H(BK).$$
On the other hand,
\begin{align}
H(Y)&=H(Y|M)+I(Y;M)\notag\\
&\geq H(Y|M)\notag\\
&= H(AM+BK|M)\notag\\
&= H(BK|M)\notag\\
&= H(BK), \notag
\end{align}
which means that $H(Y)\geq H(BK)$ and the
equality holds if and only if $I(Y;M)=0$.
\end{IEEEproof}

\begin{IEEEbiographynophoto}{\bf Fan Cheng}(S'12-M'14)
received the bachelor degree in
computer science from Shanghai Jiao Tong
University in 2007, and the PhD degree in
information engineering from The Chinese
University of Hong Kong in 2012. As of 2012,
he has been a postdoctoral fellow in the Institute of
Network Coding.
\end{IEEEbiographynophoto}

\begin{IEEEbiographynophoto}
{\bf Raymond W. Yeung} (S'85-M'88-SM'92-F'03)
was born in Hong Kong on June 3, 1962.  He received the
B.S., M.Eng., and Ph.D.\ degrees in electrical engineering from Cornell
University, Ithaca, NY, in 1984, 1985, and 1988, respectively.

He was on leave at Ecole Nationale Sup\'{e}rieure des
T\'{e}l\'{e}communications, Paris, France, during fall 1986.  He
was a Member of Technical Staff of AT\&T Bell Laboratories
from 1988 to 1991.
Since 1991, he has been with
The Chinese University of Hong Kong, where he is now
Choh-Ming Li Professor of Information Engineering and Co-Director of Institute of Network Coding.
He is also a Changjiang Chair Professor at Xidian University (2009-12)
and an Advisory Professor at Beijing University of Post and Telecommunications
(2008-11).
He has held visiting positions
at Cornell University, Nankai University, the University of
Bielefeld, the University of Copenhagen, Tokyo Institute of
Technology, and Munich University of Technology.
He was a consultant
in a project of Jet Propulsion Laboratory, Pasadena, CA, for salvaging
the malfunctioning Galileo Spacecraft and a consultant for NEC, USA.

His research interests include information theory and network coding.
He is the author of the textbooks
{\em A First Course in Information Theory}
(Kluwer Academic/Plenum 2002) and its revision
{\em Information Theory and Network Coding} (Springer 2008), which have
been adopted by over 60 institutions around the world.
In spring 2014, he gave an MOOC on information theory on Coursera based on his second book.
This book has also been published in Chinese (Higher Education Press 2011, translation by
Ning Cai~{\em et~al.}).

Dr.\ Yeung was a member of the Board of Governors of the IEEE Information
Theory Society from 1999 to 2001.
He has served on the committees of a number of information theory symposiums
and workshops.  He was General Chair of the
First and the Fourth Workshops on Network, Coding, and Applications (NetCod 2005
and 2008),
a Technical Co-Chair for the 2006 IEEE
International Symposium on Information Theory, and a Technical Co-Chair
for the 2006 IEEE Information Theory Workshop, Chengdu, China.
He currently serves as an Editor-at-Large
of {\em Communications in Information and Systems},
an Editor of {\em Foundation and Trends in Communications and Information
Theory} and of {\em Foundation and Trends in Networking},
and was an Associate Editor for Shannon Theory of the {\em IEEE Transactions
on Information Theory} from 2003 to 2005.
In 2011-12, he serves as a Distinguished Lecturer of the IEEE Information Theory Society.

He was a recipient of the Croucher Foundation Senior
Research Fellowship for 2000/2001,
the Best Paper Award (Communication Theory) of the
2004 International Conference on Communications, Circuits
and System (with C.~K.~Ngai),
the 2005 IEEE Information Theory Society Paper Award
(for his paper ``Linear network coding'' co-authored with S.-Y.~R.~Li
and N.~Cai), and the Friedrich Wilhelm Bessel Research Award of the Alexander von Humboldt Foundation in 2007.
He is a Fellow of the IEEE and the Hong Kong Institution of Engineers.
\end{IEEEbiographynophoto}
\end{document}